\documentclass[aps,prb,twocolumn,amsmath,amssymb,superscriptaddress,citeautoscript]{revtex4-2}
\usepackage{graphicx}% Include figure files
\usepackage{dcolumn}% Align table columns on decimal point
\usepackage{bm}% bold math
\usepackage{latexsym,epsfig}
\usepackage{graphicx}
\usepackage[export]{adjustbox}
\usepackage{verbatim}
\usepackage{comment}
\usepackage{amsmath}
\usepackage{amssymb}
\usepackage{stmaryrd}
\usepackage{color}
\usepackage{epstopdf}
\usepackage{grffile}
\usepackage{mathtools}
\usepackage{physics}
\usepackage{ulem}
\DeclareGraphicsExtensions{.eps}

\usepackage{hyperref}
\hypersetup{
    colorlinks=true,      
    urlcolor=blue,
    citecolor=blue,
    linkcolor=blue
}

\begin{document}
\title{Delocalization in a partially disordered interacting many-body system}
\author{Suman Mondal}
\affiliation{Institut f\"{u}r Theoretische Physik, Georg-August-Universit\"{a}t G\"{o}ttingen, D-37077 G\"{o}ttingen, Germany}
\author{Fabian Heidrich-Meisner}
\affiliation{Institut f\"{u}r Theoretische Physik, Georg-August-Universit\"{a}t G\"{o}ttingen, D-37077 G\"{o}ttingen, Germany}

\date{Mar. 13, 2024}

\begin{abstract}
We study a partially disordered one-dimensional system with interacting particles. Concretely, we  impose a disorder potential to only every other site, followed by a clean site. Our numerical analysis of eigenstate properties is based on the entanglement entropy and density distributions.
Most importantly, at large disorder, there exist eigenstates with large entanglement entropies and significant correlations between the clean sites. 
These states have  volume-law scaling, embedded into a sea of area-law states, reminiscent of inverted quantum-scar states. 
These eigenstate features leave fingerprints in the nonequilibrium dynamics  even in the large-disorder regime, with a strong
initial-state dependence. 
We demonstrate that certain types of initial charge-density-wave states decay significantly, while others preserve their initial
inhomogeneity, the latter being the typical behavior for many-body localized systems.
 This initial-condition dependent dynamics may give extra control over the delocalization dynamics at large disorder strength and should be experimentally feasible with ultracold atoms in optical lattices.
\end{abstract}

\maketitle
\section{Introduction}
A generic quantum many-body system is expected to thermalize through its intrinsic dynamics~\cite{dAlessio2016,Gogolin2016,Deutsch2018,Mori2018}. There are exceptional cases where the system refuses to thermalize, such as integrable systems~\cite{Rigol2007,Iucci2009,Polkovnikov2011,Calabrese2011,Essler2013,Vidmar2016}, which are fine-tuned systems. The many-body localized (MBL) phase, a phenomenon in which a quantum-mechanical many-body system ceases to thermalize in the presence of disorder~\cite{Nandkishore2015,Altman2018,Abanin2019,Alet2018}, is a generic counter-example of a thermalizing system. However, there are different viewpoints on whether a stable MBL phase exists or not in the thermodynamical limit.

On the one hand, numerical evidence in small systems implies many-body localization in the presence of large random potentials~\cite{Oganesyan2007,Znidaric2008,Pal2010,Bardarson2012,Luitz2015,BarLev2015,Bera2015} (for other theoretical arguments in favor of MBL, see \cite{Basko2006,Gornyi2005,Imbrie2016} and review articles \cite{Nandkishore2015,Altman2015,Abanin2019}). 
Experimental observations also advocate localization at large disorder, signaling clear signatures of MBL phases in finite systems and on accessible time scales~\cite{Schreiber2015,Choi2016,Roushan2017,Kai2018,Lukin2019,Aidelsburger2019,Leonard2023}. 
Recently, a large pre-thermal MBL regime has been predicted that results from many-body resonances and exhibits exponentially long thermalization times~\cite{Morningstar2022,David2023}. This pre-thermal MBL phase moves the MBL phase into a larger-disorder regime compared to the original predictions \cite{Luitz2015,Morningstar2022}.

On the other hand, the interpretation of numerical results that are in favor of MBL 
on small systems has been challenged in a series of studies altogether, even in one dimension
\cite{Suntajs2020,Panda2019,Suntajs2020a,Sels2021,LeBlond2021,Kiefer-Emmanouilidis2020,Kiefer-Emmanouilidis2021,Sierant2022,Evers2023} (see also \cite{Luitz2020,Abanin2021}). The main mechanisms proposed for
destabilizing MBL are many-body resonances (see, e.g., \cite{Morningstar2022} and references therein) and so-called avalanches \cite{Wojciech2017,Sels2021,Morningstar2022,Huse2023,Thiery2018,Luitz2017,Goihl2019}, a nonperturbative effect.
 According to that theory, a small thermal bubble can serve as a bath and ultimately cause the entire system to thermalize through the propagation mechanism of an avalanche. 
A recent experiment studies a partially disordered system to seek experimental evidence for quantum avalanches~\cite{Leonard2023}. Concretely, an interface between a disordered and a clean system is studied with ultracold atoms in optical lattices. An accelerated penetration of the thermal bath through the interface has been interpreted as evidence of  such quantum avalanches.

\begin{figure}[t]
\centering
    \includegraphics[clip, trim={{1.2\linewidth} {0\linewidth} {0.0\linewidth} {0.9\linewidth}}, width=1.\linewidth]{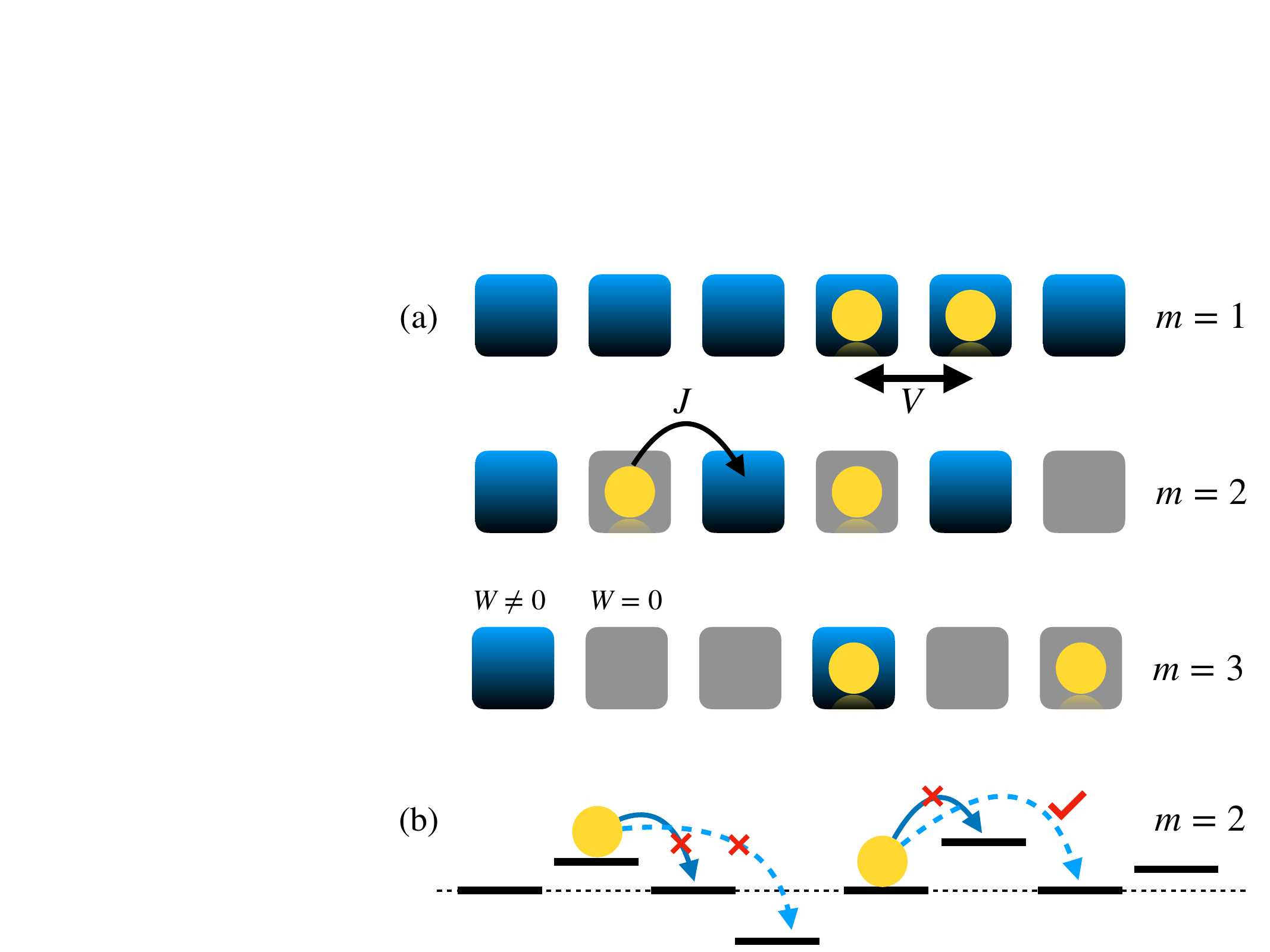}
\caption{(a) Schematic diagram of the model Hamiltonian ($\hat{H}_m$) from Eq.~(\ref{eq:ham}) shown for $m=1$, $2$ and $3$. The light gray (deep blue) sites are clean (disordered) sites, i.e., $\epsilon_i=0$ or $\epsilon_i$ random numbers. The particles (yellow circles) can hop to the nearest-neighbor site with hopping amplitude $J$ and two particles sitting at nearest-neighbor sites interact with energy $V$. (b) The effective correlated-hopping between the clean sites at large disorder strength is pictorially represented for the $m=2$ case (solid arrows are direct while dashed lines indicate higher-order hopping processes). See the discussion in Sec.~\ref{sec:diag}. 
}
\label{fig:model}
\end{figure}

Such partially disordered models can therefore provide insights into  the localization and \textit{delocalization} properties in  disordered systems. So far, several models have been studied in this context but mostly in the non-interacting limit, such as the mosaic lattice with quasi-periodic potentials~\cite{Wang2020} and a partially disordered ensemble of random regular graphs~\cite{Daniil2023}, which has mobility edges that essentially extend to infinite disorder potentials. Even though the non-interacting limits of such systems are well-studied, less is known about the effect of interactions in these kinds of systems. 

We study a particularly simple example of patterned disorder (see Fig.~\ref{fig:model}(a) for a sketch), namely a one-dimensional partially disordered system, where not all sites of the system are subject to  the disorder potential, but an equal number of disorder-free (clean) sites separate disordered sites, similar to the mosaic lattice~\cite{Wang2020}. The random disorder potential is sampled from a uniform box distribution. In the single-particle spectrum, there exist unique states that are completely extended and do not localize even for very large values of disorder strength. Therefore, we expect interesting effects to occur in the many-body case as well.

 In the interacting case with hardcore bosons, we first observe a larger ergodic regime compared to the fully disordered case on finite systems. This, however, can be understood by the reduced variance of the disorder distribution due to the presence of clean sites.
 More importantly, the patterned disorder introduces
 highly-entangled states in the background of low-entanglement states, 
 where the latter is typical for a finite-size system in the MBL regime.

 Consequently, by choosing appropriate initial states, the 
 presence of the patterned disorder affects the temporal dynamics and a strong initial-condition dependent dynamics is observed. Depending on whether particles are initially localized on disordered or clean sites, 
 there is either a strongly nonergodic dynamics
 or a significant decay of density imbalances, respectively. 
 The latter is associated with retaining less memory about the initial state on the same system sizes in the measurement of local densities and such memory is increasingly erased on larger system sizes.
 
A similar phenomenon has been found in an interacting disordered ladder system with hardcore bosons \cite{DasSarma2018}. Depending on the initial condition, that system shows thermalizing or localizing behavior. Another such  coexistence is present in  a two-dimensional interacting system~\cite{Antonio2022}.
Our results can be cast into the language of random hopping models~\cite{Eggarter1978,Soukoulis1981,Evangelou1986,Roman1987,Roman1988,Akshay2021} (see Fig.~\ref{fig:model}(b) for a sketch): The presence of the disorder sites leads to effective random hopping processes between the clean sites, while processes between disordered sites are suppressed.

Our observation of highly-entangled states in a sea of area-law states is reminiscent of the recently discussed inverted quantum-scar states \cite{Srivatsa2023,Iversen2022,Chen2023} for models different from ours (see also \cite{Srivatsa2020,Iversen2023, Iversen2023a} and \cite{Serbyn2021,Chandran2023} for reviews on quantum-scar states).
In our case, these states are not {quantum scars}, just the phenomenology with respect to entanglement is similar.
Our set-up provides an experimentally easy-to-implement scheme to introduce fast-decaying dynamics in 
strongly disordered systems and a significant initial-state dependence. Patterned disorder such as envisioned
in our work can easily be implemented in quantum-gas microscopes with digital mirror devices.

The rest of the paper is structured in the following way. We explain the model and method to simulate the partially disordered system in Sec.~\ref{sec:mod}. We explore the properties of eigenstates of finite systems with interaction in Sec.~\ref{sec:static},
where we establish the existence of eigenstates with large entanglement and significant density-density correlations between disorder-free sites in the large-disorder regime. Section ~\ref{sec:dynamics} presents the initial state-dependent relaxation dynamics. The diagonal ensemble average of different observables, such as density imbalance and density-density correlations, is provided in Sec.~\ref{sec:diag} to summarize the localization properties in such systems. Finally, we conclude in Sec.~\ref{sec:con}. An appendix contains a discussion of non-interacting systems, finite-size dependencies of the von Neumann entropy,
its time dependence, finite-size dependencies of density correlators,  and additional results for half filling.

\section{Model and method}\label{sec:mod}
The one-dimensional partially disordered model, illustrated in Fig.~\ref{fig:model}, where clean sites are periodically placed between the disordered sites, can be described by the Hamiltonian
\begin{align}\label{eq:ham}
    \hat{H}_m  = & -J\sum_{\langle i,j \rangle} \left(\hat{b}_i^\dagger \hat{b}_{j} + \rm{h.c.}\right) + \sum_{i \in [1,L,m] } \epsilon_i  \hat{n}_i \nonumber\\
    & + V\sum_{\langle i,j \rangle} \hat{n}_i \hat{n}_j.
\end{align}
Here, $\hat{b}_i^\dagger$ ($\hat{b}_i$) and $\hat{n}_i$ are bosonic creation (annihilation) and number operators at a given site $i$. The system has an onsite hardcore constraint where each site has a local Hilbert space of dimension $d=2$ ($\{|0\rangle, |1\rangle\}$) and $\hat{b}^{\dagger 2}_i |0\rangle= 0$. The first term defines the nearest-neighbor hopping process with matrix element $J$, with $\langle i,j \rangle$ indicating nearest neighbors. The second term adds disorder to the Hamiltonian. The onsite potentials $\epsilon_i$ are random numbers uniformly drawn from a box distribution $ \epsilon_i \in [-W, W]$. Here, the notation $i\in[1,L,m]$ stands for a site index $i$ that varies between $[1, L]$ with regular interval $m$. This makes the system periodically disordered where $m-1$ disorder-free sites separate two disordered sites (see Fig.~\ref{fig:model}). The third term in the Hamiltonian stands for nearest-neighbor interactions with interaction strength $V$. Note that we use periodic-boundary conditions for our calculations. Here, we consider $J=1$, defining the unit of energy and other parameters and observables. The energy density is defined as 
\begin{align}
\epsilon = \frac{E_\alpha - E_{\text{min}}}{E_{\text{max}} - E_{\text{min}}}\,,
\end{align}
where $E_\alpha$ is an eigenenergy of $\hat H_m$ and $E_{\text{min}}$ ($E_{\text{max}}$) is the groundstate (antigroundstate) energy for a given disorder configuration.

In the non-interacting limit ($V/J=0$) and for $m=1$, the model is the well-known Anderson-disorder model~\cite{Anderson1958,Ramakrishnan1985}. In this case, we know that all single-particle eigenstates localize for any finite $W$. Switching on interactions, i.e., $V>0$, the system exhibits many-body localization at large $W$ in  finite systems~(see, e.g., \cite{Oganesyan2007,Pal2010,Luitz2015,BarLev2015,Bera2015,Lim2016}). 
For $m=1$, and with our choice of units, the finite-size crossover scale to localization is at $W_c \gtrsim 4J$ (see, e.g., \cite{Luitz2015}).
For the $m=2$ case and at $V=0$, there exists an extended state in the middle of the spectrum for all disorder strength $W$ (see App.~\ref{sec:app}). We anticipate that an analogon will persist in the many-body case. To analyze the many-body system, we consider a filling of $\rho=1/4$.
There are two reasons: First, larger system sizes can be reached, second, localization is more stable at lower filling factors \cite{Hopjan2021}.
Note, though, that our main observations also apply to half filling $\rho=1/2$ as shown in App.~\ref{app:corr-D}.

Unless otherwise mentioned, we diagonalize the Hamiltonian fully to calculate all the observables from the eigenstates. To calculate the Hamiltonian, we use the QuSpin library~\cite{QuSpin1,QuSpin2}. The time evolution of the initial states is performed by exponentiating the Hamiltonian using the scaling-and-squaring algorithm implemented in SciPy's linear algebra library~\cite{SciPy}. All observables presented in the results sections are averaged over $N_r = 1000$ disorder realizations.
Our numerical data for interacting systems are computed for $V=2J$.

\begin{figure}
\centering
\includegraphics[width=1\linewidth]{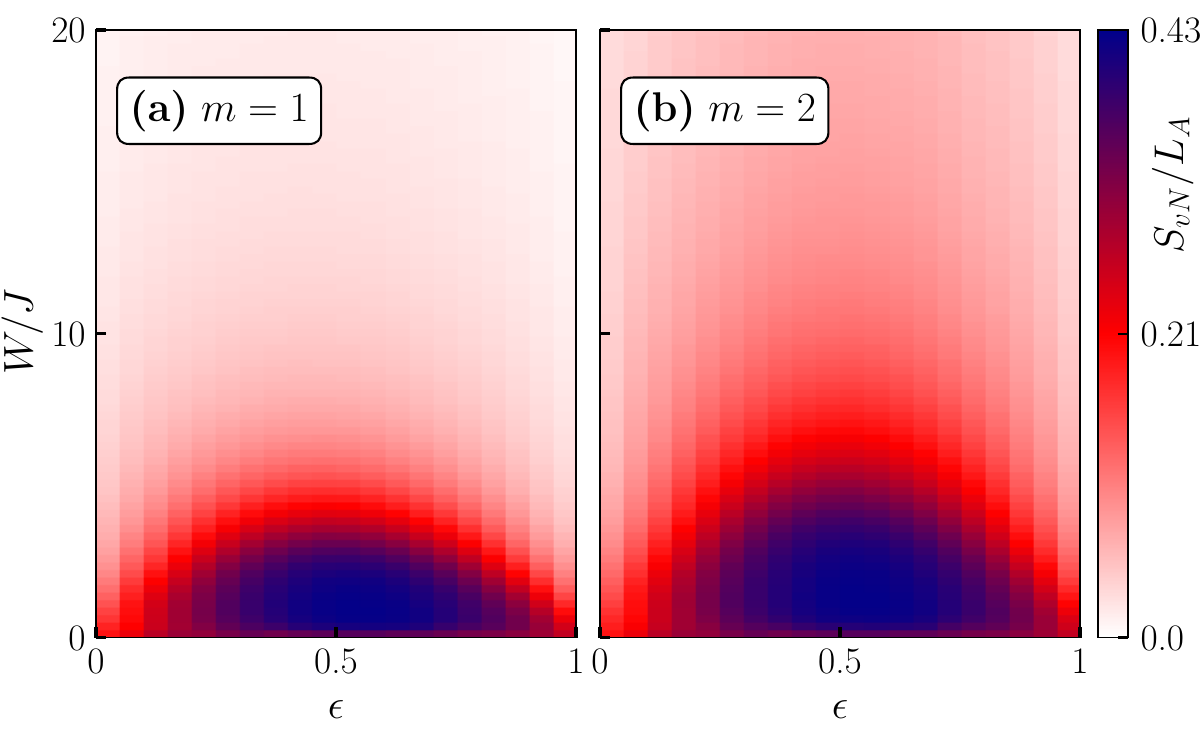}
\caption{The von Neumann entanglement entropy $S_{vN}$ is plotted in the $W$ versus energy density $\epsilon$ plane for the (a) $m=1$ and (b) $m=2$ cases with $V=2J$. Here, we consider $20$ bins in the energy density for a system of $L=16$. We see an expanded ergodic region for the $m=2$ case, which, however, can be explained by the reduced variance of the distribution of disorder potential strengths due to the presence of clean sites.}
\label{fig:pds}
\end{figure}

\section{Eigenstate properties of interacting system}\label{sec:static}
In this section, we analyze the Hamiltonian $\hat{H}_m$ in many-body setups with $V>0$. We consider a filling of $\rho = 1/4$ and calculate the entanglement entropy, density distributions, and correlations from the eigenstates. Note that in finite systems, $\hat{H}_{m=1}$ is expected to show many-body localization behavior at large $W/J$ as this system is the paradigmatic random-disorder model with nearest-neighbor interactions~\cite{Pal2010,Oganesyan2007,Luitz2015,BarLev2015,Bera2015,Lim2016}.  In the following, we study and compare the $m=2$  with the $m=1$ case to elucidate the effect of the patterned disordered potential on the many-body states.

\begin{figure}
\centering
\includegraphics[width=1\linewidth]{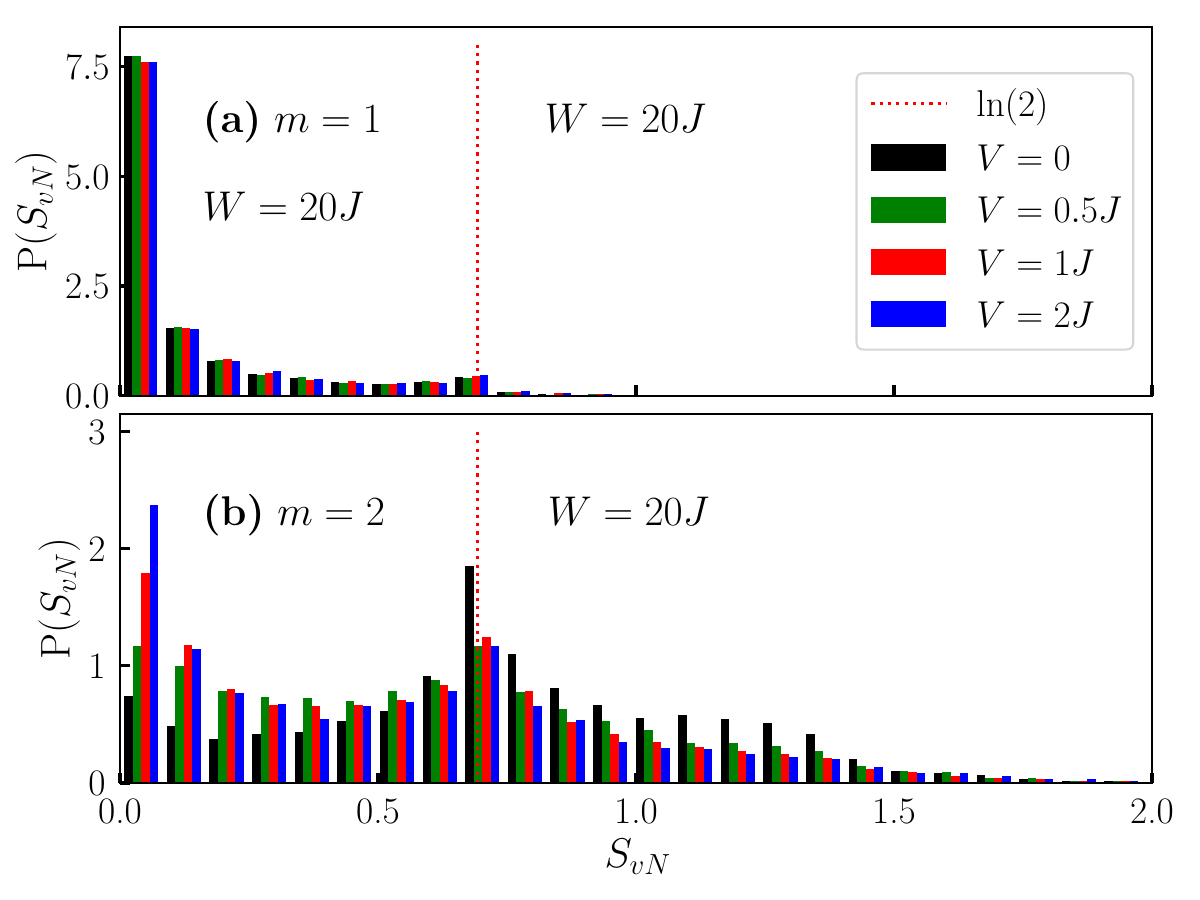}
\caption{Distribution of von Neumann entanglement entropy $S_{vN}$  
for (a) $m=1$ and (b) $m=2$ for a system of $L=16$ sites  at $\epsilon = 0.5$ (averaged over ten eigenstates). We take $W=20J$ and consider different interaction strength $V$, represented by different colors. In (a) and (b), we divide the $S_{vN}$ axis into 20 and 30 bins, respectively. The dashed lines in (a) and (b) represent $S_{vN} = {\rm{ln}}(2)$. We can see a significant peak at $S_{vN} = {\rm{ln}}(2)$ for $m=2$ compared to $m=1$ case and most importantly, the emergence of a tail of highly-entangled states.
}
\label{fig:entr_hist}
\end{figure}

\begin{figure}
\centering
\includegraphics[clip, trim={{1.7\linewidth} {0\linewidth} {0.0\linewidth} {0.0\linewidth}}, width=1\linewidth]{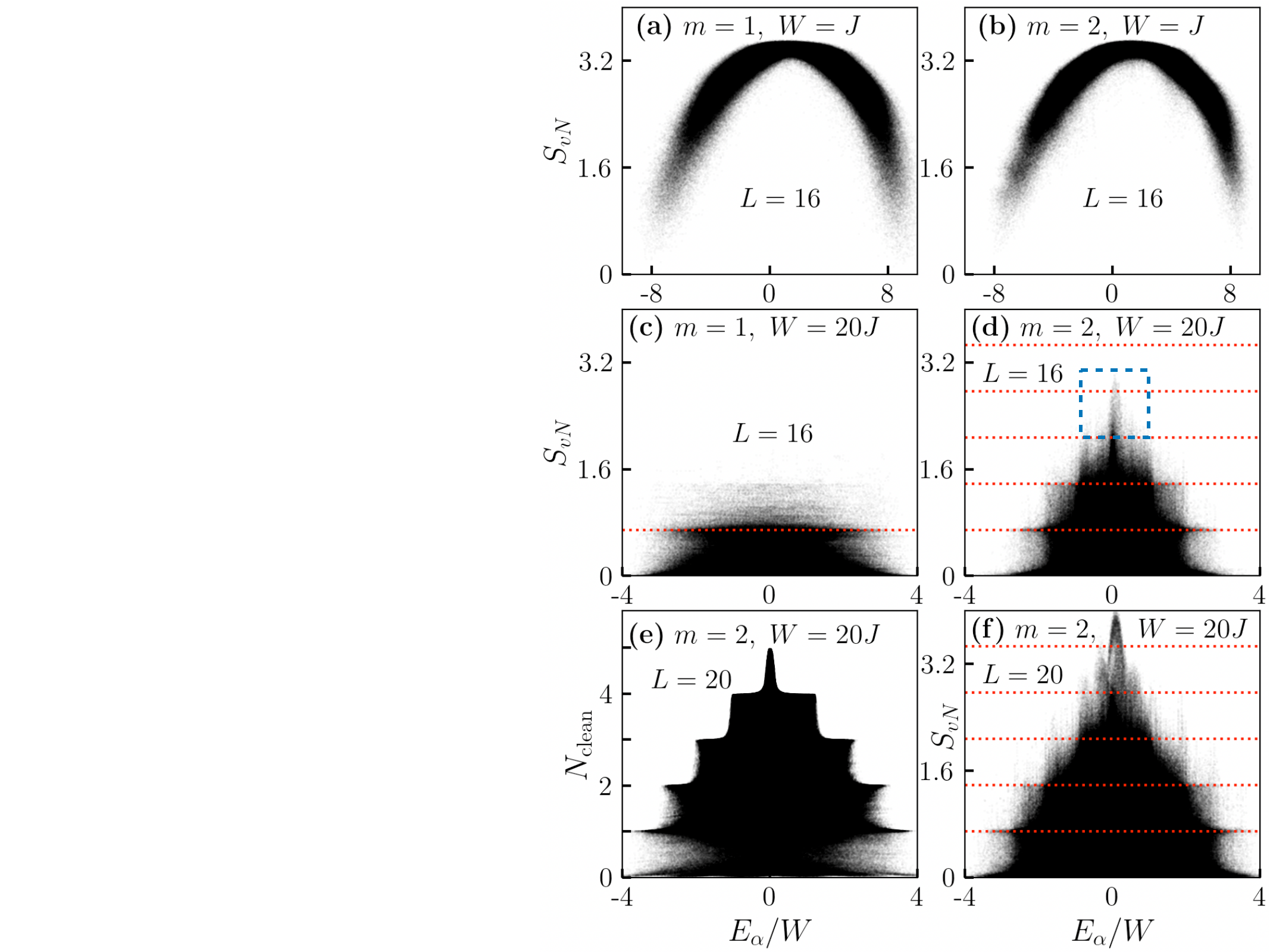}
\caption{(a)--(d): Entanglement entropy versus eigenstate energy for (a),(c) $m=1$ and
(b),(d) $m=2$ for $L=16$ sites and (a), (b) $W/J=1$ (c), (d) $W/J=20$. The box in (d) indicates the highly-entangled states. 
 (e) Number of particles in the clean sites a versus eigenstate energy for $m=2$ case with $L=20$ and $W=20J$. 
 (f) The same as (d), but for $L=20$.
 The horizontal dotted lines in (c), (d), and (f) indicate $S_{vN}=M\ln(2)$ with $M=1,2,3,4,5$.
}
\label{fig:entr_nb}
\end{figure}

\begin{figure*}[t]
\centering
\includegraphics[width=0.49\linewidth]{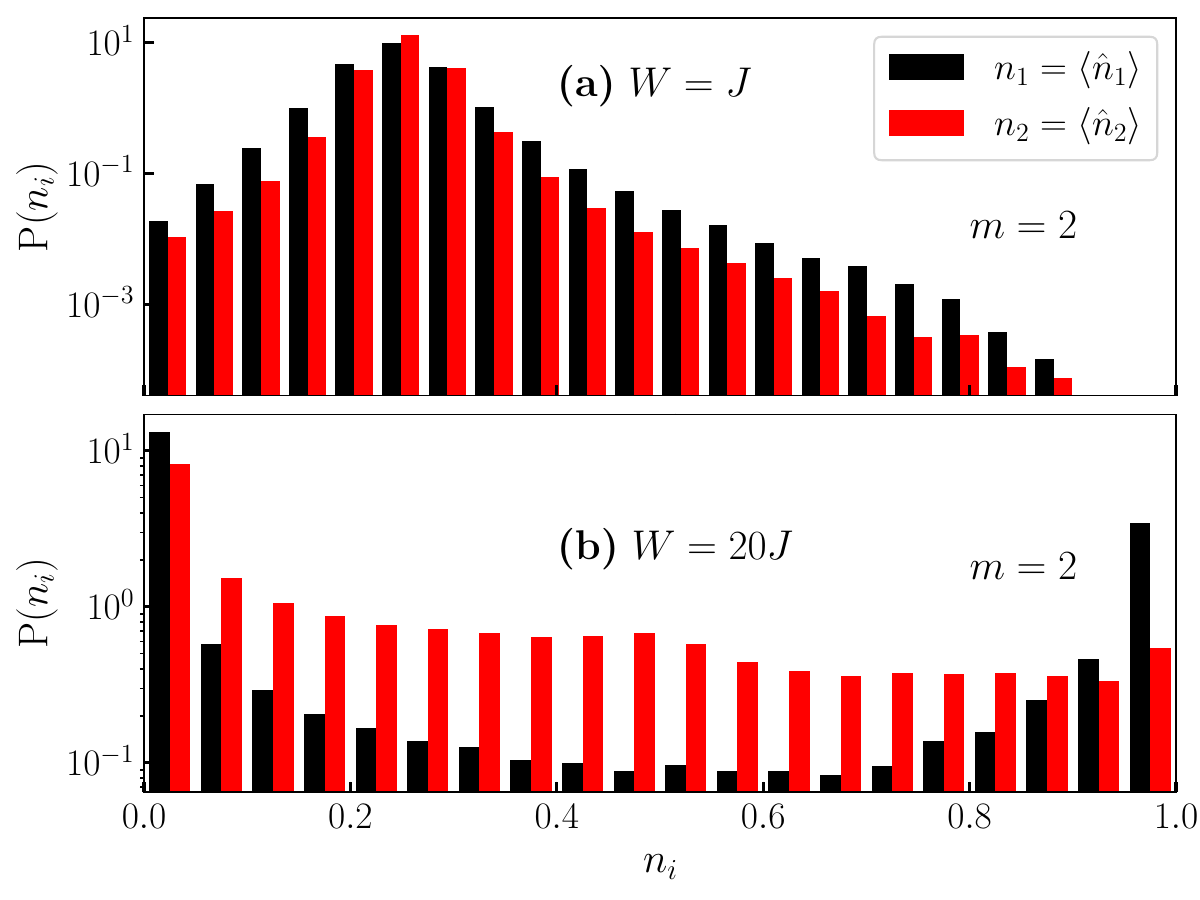}
\includegraphics[width=0.49\linewidth]{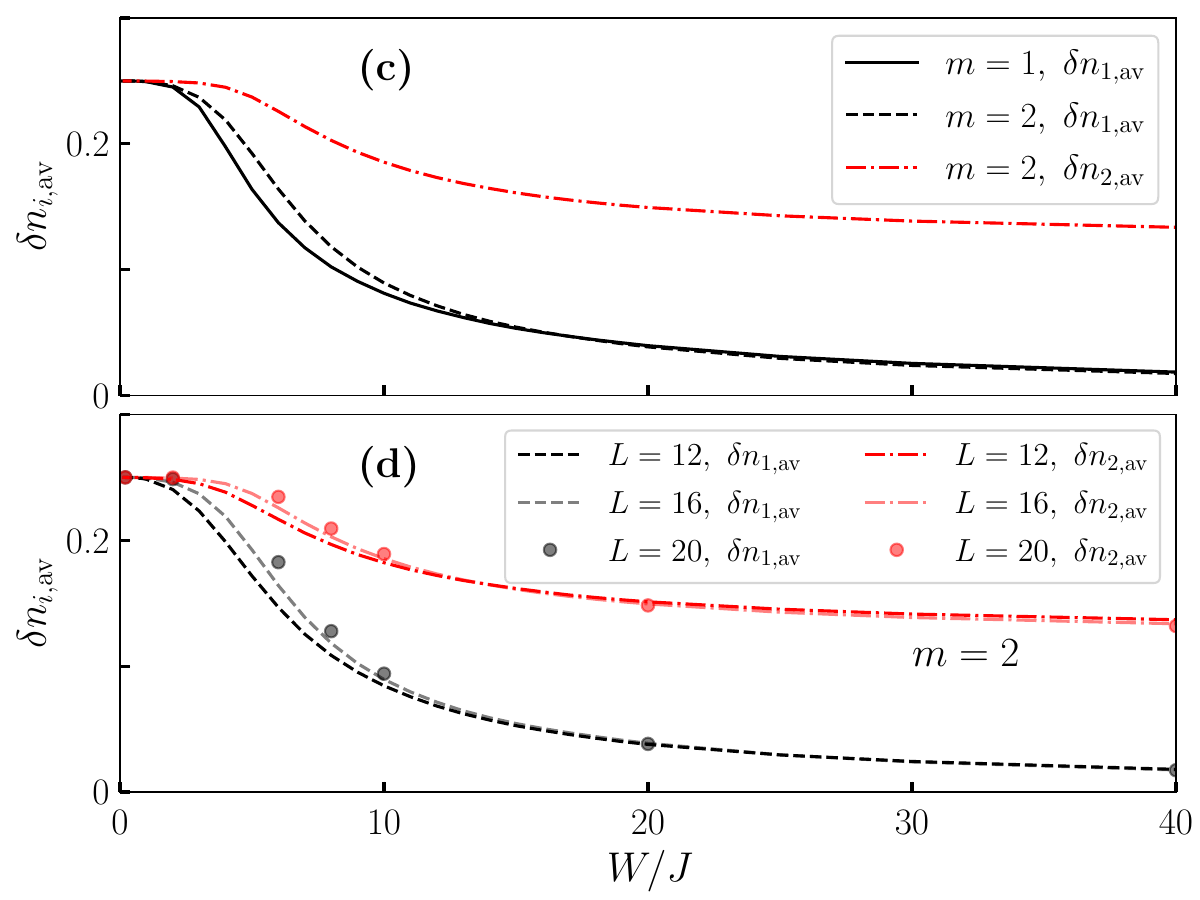}
\caption{Distributions of the onsite densities $n_i = \langle \hat{n}_i \rangle$ in eigenstates for the $m=2$ case at a site with disorder (here $i=1$) and one clean site ($i=2$), represented by black and red colors, respectively, for  (a) $W=J$ and (b) $W=20J$. At small $W=J$, the distribution peaks at the total density $\rho = 1/4$ of the system. At large disorder $W=20J$, the disordered (clean) site exhibits a bimodal (trimodal)  distribution. (c) Occupation distance $\delta n_i$ for the disordered and clean sites (the difference only matters for the $m=2$ case) for both the $m=1$ and $m=2$ case as a function of $W$. (d) Finite-size dependence of $\delta n_i$ for $m=2$.}
\label{fig:ni_hist}
\end{figure*}

\subsection{Entanglement entropy}
The bipartite entanglement entropy of the many-body eigenstates can be exploited to diagnose the degree of localization in  many-body eigenstates. Localized states are expected to follow an area law of entanglement, whereas the delocalized states are expected to follow a volume law of entanglement~\cite{Bauer2013,Kjaell2014}. We can calculate the bipartite von Neumann entanglement entropy ($S_{vN}$) of a subsystem $A$ from the reduced density matrix $\hat\rho_A$ after dividing the system into two subsystems $A$ and $B$ with length $L_A$ and $L_B$ as 
\begin{equation}
    S_{vN} = -{\rm{Tr}} \, \left[\hat \rho_A {\rm{ln}} (\hat \rho_A)\right], 
\end{equation}
where $\hat \rho_A = {\rm{Tr}}_B (|\Psi\rangle \langle \Psi|)$ is the reduced density matrix of the subsystem and $|\Psi\rangle$ is the  pure state of the whole system. Throughout, we set $L_A=L/2$. 

In Fig.~\ref{fig:pds}, we plot $S_{vN}/L_A$ with respect to energy-density $\epsilon$ and $W/J$  computed in the eigenstates of a system of $L=16$ for a finite interaction strength  $V=2J$. Figures~\ref{fig:pds}(a) and (b) contain results for the $m=1$ and $2$ cases, respectively. Compared to the $m=1$ case,  a larger region is occupied by the ergodic phase in the $m=2$ case, evidenced by large values of $S_{vN}$. 
This observation can simply be explained by the fact that
the standard deviation of disorder potentials for the $m=2$ case is a factor of  $\frac{1}{\sqrt{2}}$ smaller than the one for the $m=1$ case due to the presence of clean sites in the former case. One can rescale the $W/J$ axis of Fig.~\ref{fig:pds}(b) by $\frac{1}{\sqrt{2}}$ and compare to Fig.~\ref{fig:pds}(a), resulting in a comparable  extension of the ergodic regime in both the $m=2$ and $m=1$ cases [not shown here].
The finite-size dependence of the entanglement entropy is discussed in  App.~\ref{app:svn-L}.
As we shall see next, the key effect of the patterned disorder is to introduce a set of highly entangled states.

Investigating further, we find that the $m=2$ case gives rise to the emergence of a tail of high-entanglement states and at the same time, an increase in the number of two-particle resonances in the eigenstates, which is reflected in the distribution of $S_{vN}$ and a peak at $\ln(2)$. 
In Figs.~\ref{fig:entr_hist}(a) and (b), we plot this distribution for the $m=1$ and $2$ cases, respectively, at $\epsilon = 0.5$ for different interaction strengths and at a large disorder strength $W=20J$. 
For $m=1$  and for all values of $V$ considered here, the distribution resembles the typical distribution of an MBL regime with a large peak near zero and a small peak at $S_{vN} = {\rm{ln}}(2)$, the latter  a result of rarely occurring two-site resonances~\cite{Bauer2013,Elliott2015,Lim2016}. In contrast, $m=2$ leads to a different scenario. The peak at ${\rm{ln}}(2)$ followed by a tail at large $S_{vN}$ is significant compared to the peak at zero. The emergence of highly-entangled states is due to many-body resonances resulting from the many clean sites in the system.

Finite values of $V$ make the peak near zero larger compared to $V=0$, signaling a higher degree of localization, and the peak at ${S_{vN} = \rm{ln}}(2)$ and the extent of the tail decrease but remain large in comparison. Such a  distribution of $S_{vN}$ in the $m=2$ case also explains the higher $S_{vN}$ at large $W$ in Fig.~\ref{fig:pds}(b) compared to the $m=1$ case shown in Fig.~\ref{fig:pds}(a).
Nonetheless, even with interactions, there is a sizable tail at large
values of $S_{vN}$ compared to the $m=1$ case, which will also
give rise to initial-state dependent faster relaxation dynamics.

A central result of our paper is obtained from
plotting the half-chain entanglement entropy versus eigenstate energy, shown in Figs.~\ref{fig:entr_nb}(a)-(d) for a system of $L=16$. At weak disorder [$W/J=1 $, Fig.~\ref{fig:entr_nb}(a),(b)], a typical band emerges, with high-entanglement states in the bulk of the spectrum, with no discernible difference between the $m=1$
and $m=2$ cases. At large disorder and for $m=1$, there are mostly low-entanglement states (spreading over the entire many-body bandwidth) and states at $S_{vN} \approx \ln(2)$ [see Fig.~\ref{fig:entr_nb}(c)]. For $m=2$, 
a significant structure emerges as is evident from 
Fig.~\ref{fig:entr_nb}(d). First, there are subsequent bands at $M \ln(2)$ ($M=1,2,3,4$), with a varying bandwidth indicated 
by the dashed horizontal lines. Second, at $E_\alpha=0$, highly-entangled states emerge that reach almost as large values 
as for weak disorder [see the box in Fig.~\ref{fig:entr_nb}(d)].
For orientation, our estimate for the maximum possible entanglement for
the clean part alone is $S_{vN}=\ln(2^M)  =  M \ln(2) \leq 2.77$ (for $M=4$)
for the parameters of Fig.~\ref{fig:entr_nb}(d), which is consistent with the data (for a discussion of $S_{vN}$ in eigenstates, see \cite{Page1993,Vidmar2017,DeTomasi2020}).
We will argue that these high-entanglement states are responsible for the initial-state dependent fast relaxation dynamics discussed in Sec.~\ref{sec:dynamics}.

The highly-entangled states follow volume-law scaling as  expected from the discussion above. In  Fig.~\ref{fig:entr_nb}(f), we present $L=20$
data, to be compared with the for $L=16$ data from Fig.~\ref{fig:entr_nb}(d). We can see that there is an increment in the number of bands ($M~{\rm{ln}}(2)$) compared to the $L=16$ case. Now, the estimate for the entanglement from the clean sites alone reaches $5~{\rm{ln}}(2)\approx 3.47$, signifying the volume-law scaling. The actual $S_{vN}$ exceeds this value due to contributions from the disordered sites. The phenomenology emerging here is similar to the
case studied in \cite{Chen2023}, where another potential is introduced that localizes some but not all states.

The structure for $m=2$ in the $W\gg J$ regime can be understood by assuming different energy-dependent distributions of particles on clean and disordered sites and approximating the total entanglement $S_{vN} \approx S_{vN}^{\rm{dis}}+ S_{vN}^{\rm{clean}}$ by the sum of contributions from particles on disordered and clean sites. While there is a very minor contribution from the disordered sites in the $W\gg J$ limit, the main contribution comes from the clean sites. In Fig.~\ref{fig:entr_nb}(e), we portray the total number of particles available in the clean sites ($N_{\rm{clean}}$) versus the eigenstates. We can see the emergence of the bands with different numbers of particles at different energies. The structure in $N_{\rm{clean}}$ correlates with  the bands in $S_{vN}=M~{\rm{ln}}(2)$ with $M=N_{\rm{clean}}$.

\subsection{Density Distribution}
Some other markers which are experimentally observable can give insight into the degree of localization. In the following, we  analyze the onsite densities and density-density correlations, which may tell us if there is any distinction between clean and disordered sites for the $m=2$ case. We first look at the probability distributions ${\rm{P}}(n_i)$ of densities $n_i = \langle \hat{n}_i \rangle$ in one of the disordered sites ($i=1$) and one of the clean sites ($i=2$), computed in all eigenstates. 

Figures~\ref{fig:ni_hist}(a) and (b) show the density distributions for $W=J$ and $20J$, respectively, in a disordered and a clean site. In the small-disorder regime ($W=J$), we observe maxima in the distributions at $n_i = 1/4$, which is equal to the filling. This is the expected behavior in the ergodic regime~\cite{Hopjan2021,Hopjan2020}. At large disorder ($W=20J$), we get a clear differentiation of the distribution in clean and disordered sites. While the disordered site exhibits a bimodal distribution implying strong localization, the clean site shows a trimodal distribution. This behavior of the density distribution on clean sites deviates from the case of strong localization.
The central peak for $m=2$ is at $n\approx 0.5$, corresponding to the situation where all particles sit on the clean sites.

We can quantify this difference between the clean and disordered sites in the $m=2$ case by calculating the occupation distance $\delta n_i = |n_i - [n_i]|$ where $[n_i]$ is the closest integer of $n_i$ and study it as a function of $W$ \cite{Hopjan2021,Hopjan2020}. We plot  $\delta n_i$, averaged over eigenstates and disorder realizations ($\delta n_{i,\rm{av}}$), for disordered ($\delta n_{1,\rm{av}}$) and clean ($\delta n_{2,\rm{av}}$) sites in Fig.~\ref{fig:ni_hist}(c) for both the $m=1$ and $m=2$ cases. In the ergodic and strongly localized regime, we expect  $\delta n_{i,\rm{av}} \to \rho$ or to be small and $L$-independent, respectively. As we can see from Fig.~\ref{fig:ni_hist}(c), in the $m=1$ case, $\delta n_{i,\rm{av}}$ asymptotically goes to zero at large $W$ consistent with localization. However, for the $m=2$ case, we find differences between clean and disordered sites. The occupation distance $\delta n_{i,\rm{av}}$ for a disordered site behaves similarly to the $m=1$ case but $\delta n_{i,\rm{av}}$ for a clean site never goes close to zero even at very large $W$.  The actual  values are lower than $\rho=1/4$, the latter the  expected result in the ergodic regime, yet clearly different from the typical behavior of 
the localized case \cite{Hopjan2021}.
By increasing system size, for $W \lesssim 10J$, there is a clear trend for $\delta n_i$ to increase with $L$ towards 0.25 {[see Fig.~\ref{fig:ni_hist}(d)]}.

\subsection{Density Correlations}
\begin{figure}[t]
\centering
\includegraphics[width=1.\linewidth]{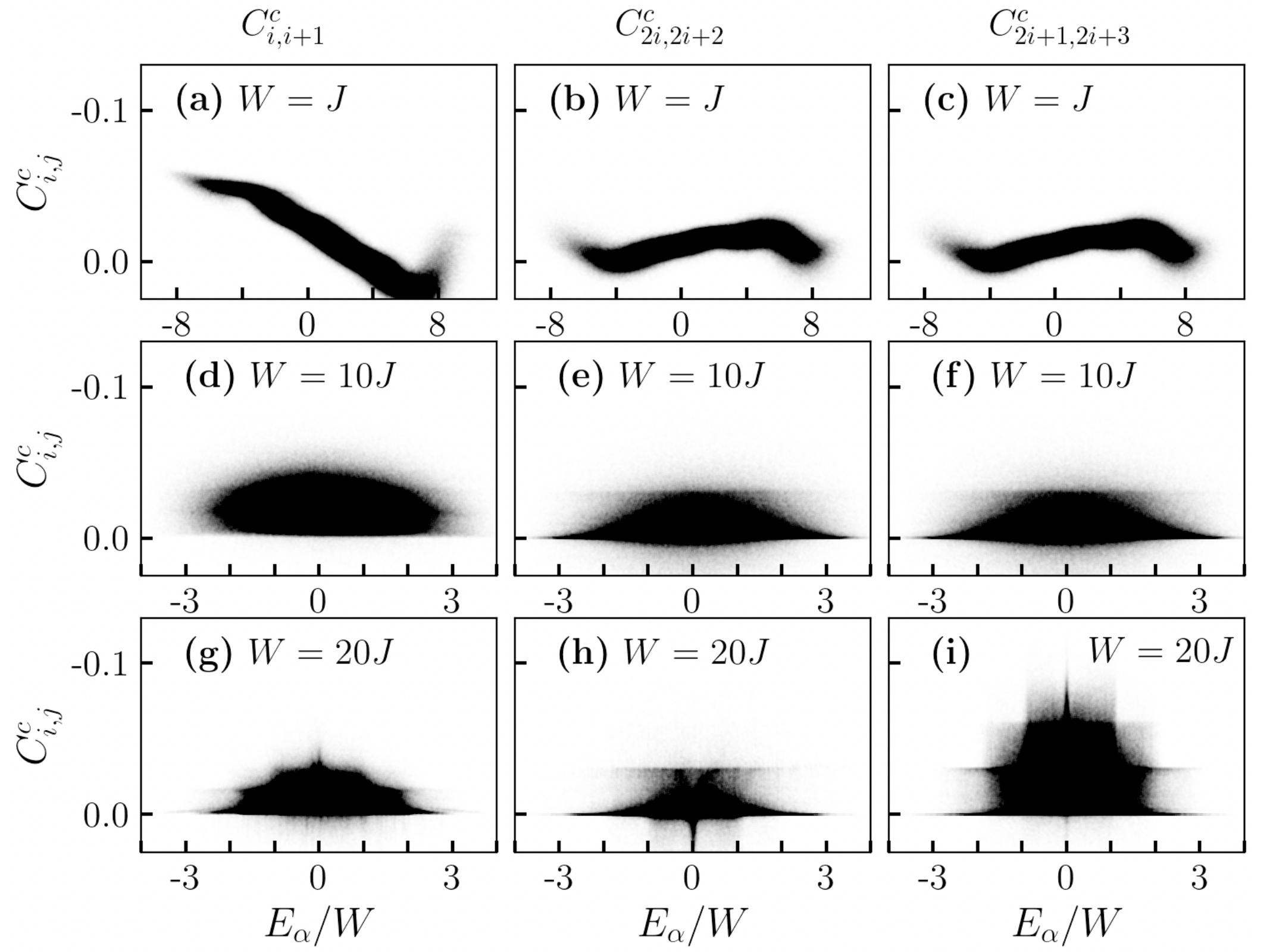}
\caption{Connected two-site density-density  correlators $C_{i,j}^c$  for (a-c) $W=J$, (d-f) $W=10J$, and (g-i) $W=20J$ for the $m=2$ case. We show the nearest-neighbor correlations between a disordered and a clean site ($C_{i,i+1}^c$), between two disordered sites ($C_{2i,2i+2}^c$), and between two clean sites ($C_{2i+1,2i+3}^c$) in the first, second, and third columns, respectively. As the disorder strength $W$ increases, the correlators $C_{i,i+1}^c$ and $C_{2i,2i+2}^c$ decrease (compare the $W=10J$ and $20J$ cases) but  $C_{2i+1,2i+3}^c$ remains large. }
\label{fig:eig_corr}
\end{figure}

Now, we analyze the nearest-distance connected density-density correlations between a disordered and a clean site ($C_{i,i+1}^c$), two disordered sites ($C_{2i,2i+2}^c$), and two clean sites ($C_{2i+1,2i+3}^c$).
The definitions are:
\begin{equation}\label{eq:corr}
    C_{ij}^c = \langle \hat{n}_i \hat{n}_j \rangle - \langle \hat{n}_i \rangle \langle \hat{n}_j \rangle\,.
\end{equation}
We plot these correlators in Fig.~\ref{fig:eig_corr} 
as a function of the eigenenergy $E_\alpha$  for the $m=2$ case, 
after averaging over the whole system, for $W=J$, $10J$, and $20J$.

In the ergodic regime ($W=J$), the correlations are smooth functions of energy with a small spreading due to the small system size. The spreading decreases with increasing system size (see App.~\ref{app:corr-C}). In the large $W$ regime, we see a non-ETH-like behavior  on finite systems \cite{Rigol2008,Luitz2016}: the correlations spread over large values at the same energy and do not shrink with increasing system size (see App.~\ref{app:corr-C}). 
Importantly,  $C_{i,i+1}^c$ and $C_{2i,2i+2}^c$ keep decreasing with increasing $W$ (compare the $W=10J$ and $20J$ cases). However, the correlation between the clean sites ($C_{2i+1,2i+3}^c$) remains finite and large in this regime (compare the $W=20J$ case). These features are absent in the $m=1$ case where in the large $W$ regime, all the correlations systematically decrease with increasing $W$.

\begin{figure}[t]
\centering
\includegraphics[clip, trim={{0\linewidth} {0\linewidth} {0.0\linewidth} {2\linewidth}}, width=0.8\linewidth]{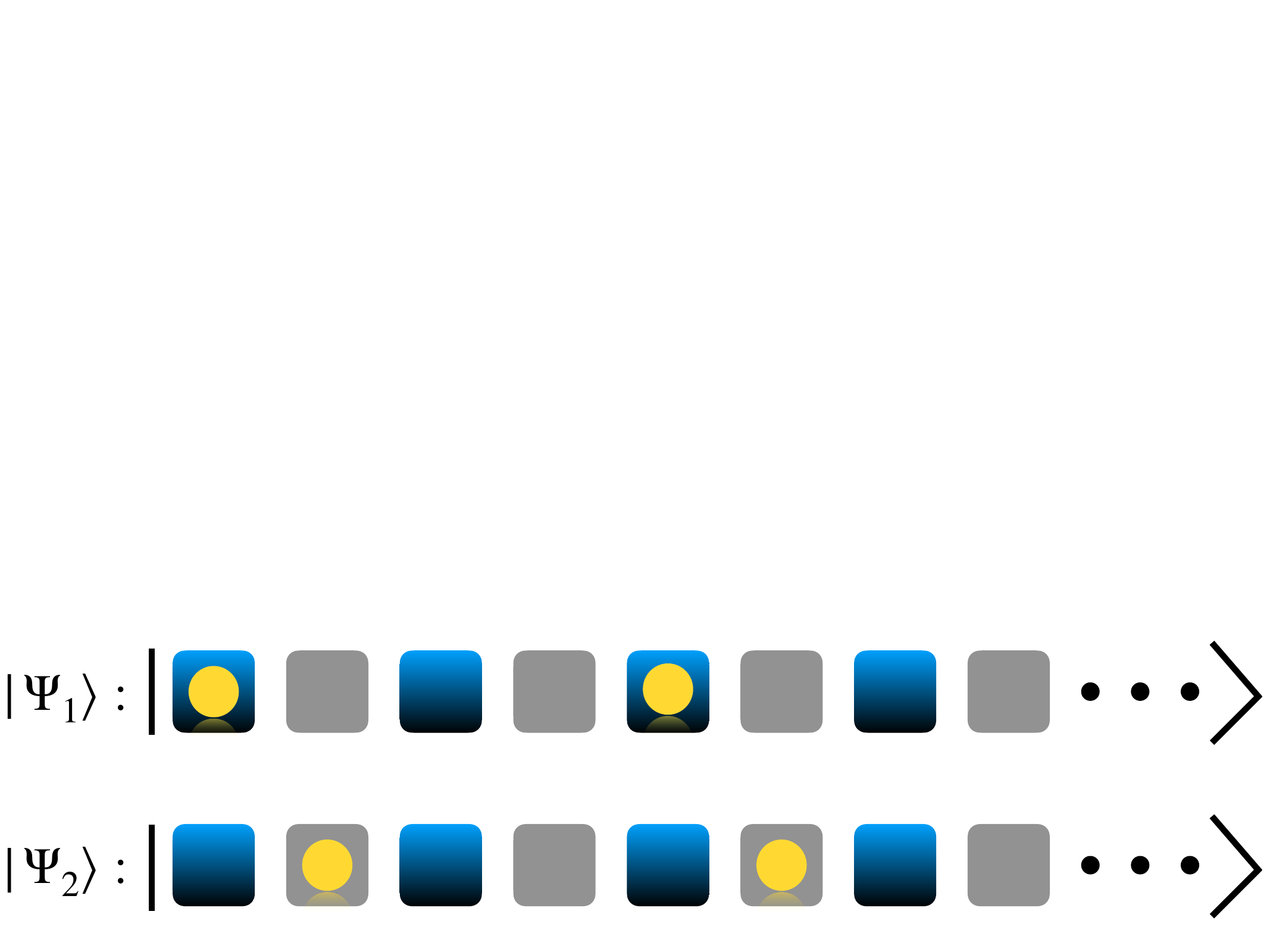}
\includegraphics[width=1\linewidth]{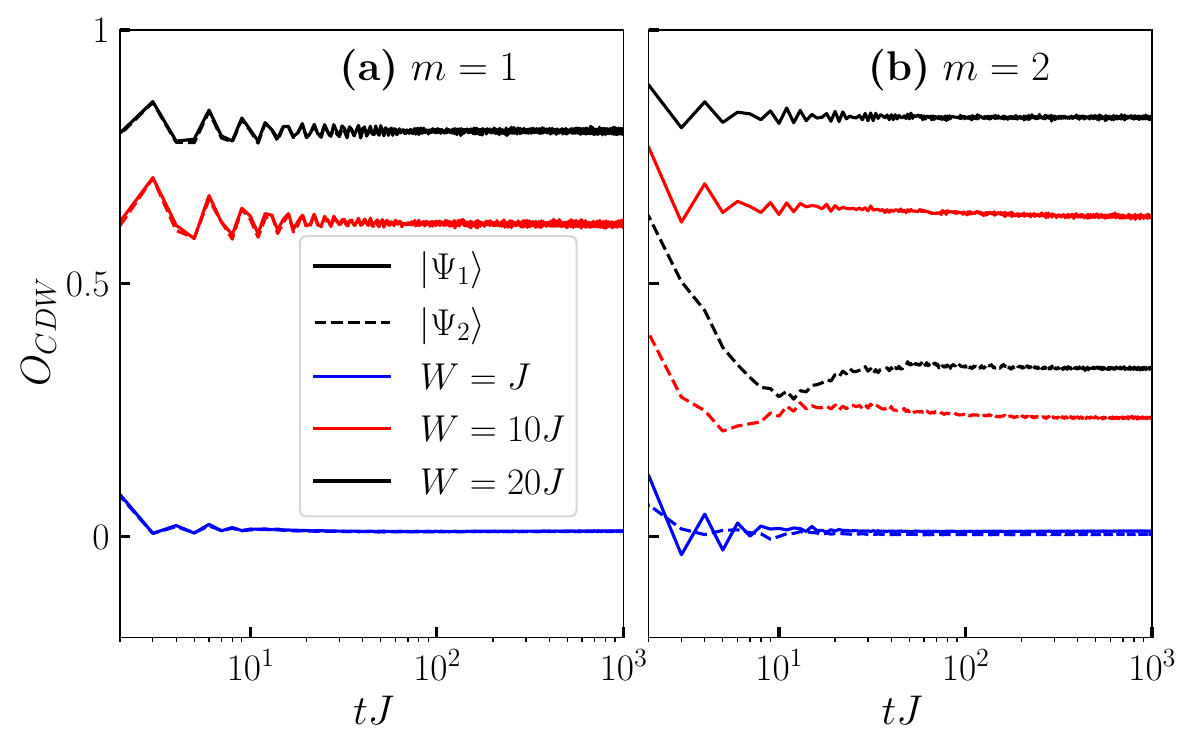}
\caption{Time dependence of the CDW order parameter $O_{CDW}$. The sketches illustrate the  dynamics for the two initial states $|\Psi_1\rangle$ and $|\Psi_2\rangle $ from Eqs.~\eqref{eq:psi1} and \eqref{eq:psi2} (see also Fig.~\ref{fig:model}). 
(a),(b): $O_{CDW}(t)$ for $L=16$ sites and $m=1$ and $2$, respectively.  Different disorder strength $W$ are represented by different colors and the two initial states $|\Psi_1\rangle$ and $|\Psi_2\rangle$ are represented by the solid and dashed lines, respectively. We can see a drastic difference in the long-time value of $O_{CDW}$ between the two initial states for the $m=2$ case at large disorder ($W=10J$ and $20J$). 
}
\label{fig:Ocdwt}
\end{figure}

One can notice a wedding cake-like structure in the correlation between the clean sites $C_{2i+1,2i+3}^c$ for large $W$ [Fig.~\ref{fig:eig_corr}(i)] which is also present in the entanglement entropy [Fig.~\ref{fig:entr_hist}(d)]. This interesting structure can be understood from the distributions of the particles in clean and disordered sites for a given eigenstate. If all the particles are in the clean sites,  $C_{2i+1,2i+3}^c$ is at its maximum and sharply peaked near $E_\alpha=0$. Now, if only one particle belongs to the disordered sites, $C_{2i+1,2i+3}^c$ decreases, and the spread over the energy window $E_\alpha \in [-W, W]$ forms the topmost step. Similarly, with an increasing number of particles in the disordered site (say $N_{\rm{dis}}$  particles out of $N$), the correlations between particles in  the clean sites decreases and form the lower steps one by one with energy windows $E_\alpha\in [-N_{\rm{dis}}W ,N_{\rm{dis}}W]$. Note that if all the particles are in the disordered sites,  $C_{2i+1,2i+3}^c$ is the lowest and spreads over the whole energy spectrum. Therefore, there is an emulsion of different classes of eigenstates depending on the number of particles in the disordered site in the many-body spectrum.
 
We will next argue that the features at zero energy -- existence of highly entangled eigenstates and large correlations between clean sites -- impact the time evolution form appropriately chosen initial states.

\section{Quench dynamics of CDW states}\label{sec:dynamics}

We investigate the quench dynamics for different initial conditions to get further insights into the system. We consider two representative initial states $|\psi(t=0)\rangle$  for the $m=2$ case, where in one case, the particles are initialized in disordered sites 
\begin{align}\label{eq:psi1}
|\Psi_1\rangle = \prod_{i=0}^{L/4-1} \hat b^\dagger_{4i+1} |0\rangle
\end{align}
and in the other case, particles are initiated from clean sites 
\begin{align}\label{eq:psi2}
|\Psi_2\rangle =\prod_{i=0}^{L/4-1} \hat b^\dagger_{4i+2} |0\rangle
\end{align}
(see the upper panel of Fig.~\ref{fig:Ocdwt}). Note that these two charge-density-wave (CDW) states are equivalent in the $m=1$ case since there, all sites are disordered. Now, we define a CDW order parameter as
(similar to the imbalance \cite{Schreiber2015})
% \begin{equation}
%     O_{CDW} =\frac{1}{N}  \sum_{i=1}^{L/4}
%     \begin{cases}
%      \langle \hat{n}_{4i} - \frac{1}{3}\hat{n}_{4i+1} - \frac{1}{3}\hat{n}_{4i+2} - \frac{1}{3}\hat{n}_{4i+3}\rangle \\ \text{if } |\psi(t=0)\rangle = |\Psi_1\rangle\\ \\
%     \langle - \frac{1}{3}\hat{n}_{4i} + \hat{n}_{4i+1} - \frac{1}{3}\hat{n}_{4i+2} - \frac{1}{3}\hat{n}_{4i+3}\rangle \\ \text{if } |\psi(t=0)\rangle = |\Psi_2\rangle \,.
%     \end{cases}
% \end{equation}
\begin{equation}
    O_{CDW} =\frac{1}{N}  \sum_{i=1}^{L/4} \langle \hat{n}_{4i} - \frac{1}{3}\hat{n}_{4i+1} - \frac{1}{3}\hat{n}_{4i+2} - \frac{1}{3}\hat{n}_{4i+3}\rangle
\end{equation}
for $|\psi(t=0)\rangle = |\Psi_1\rangle$, and
\begin{equation}
    O_{CDW} =\frac{1}{N}  \sum_{i=1}^{L/4} \langle - \frac{1}{3}\hat{n}_{4i} + \hat{n}_{4i+1} - \frac{1}{3}\hat{n}_{4i+2} - \frac{1}{3}\hat{n}_{4i+3}\rangle
\end{equation}
for $|\psi(t=0)\rangle = |\Psi_2\rangle$.

The relaxation of the CDW states is presented in Fig.~\ref{fig:Ocdwt} for a system of $L=16$ with $V=2J$. Figures~\ref{fig:Ocdwt}(a) and (b) show the evolution of $O_{CDW}$ with time for different values of $W$ (represented by different colored lines) and the two initial states (represented by two different line styles) for the $m=1$ and $2$ cases, respectively. We see the expected results for the $m=1$ case; here, the curves for both  initial states are on top of each other. At small disorder ($W=J$), there is a complete relaxation of the particle inhomgeneity, consistent with  delocalization and thermalization \cite{Luitz2016a,Sierant2022}. For large disorder $20J$, $O_{CDW}$ saturates at  a finite value, signaling localization in finite $L$~\cite{Luitz2016a,Sierant2022}. For $m=2$, we find an interesting contrast between the relaxation of the two initial states. While small disorder such as $W =J$ leads to complete relaxation for both states, for large $W$, the states relax to vastly different values of $O_{CDW}$ at a long time. We demonstrate that a similar initial-state dependent dynamics is present for  $\rho=1/2$, which is shown in App.~\ref{app:corr-D}.

\begin{figure}[t]
\centering
\includegraphics[width=1\linewidth]{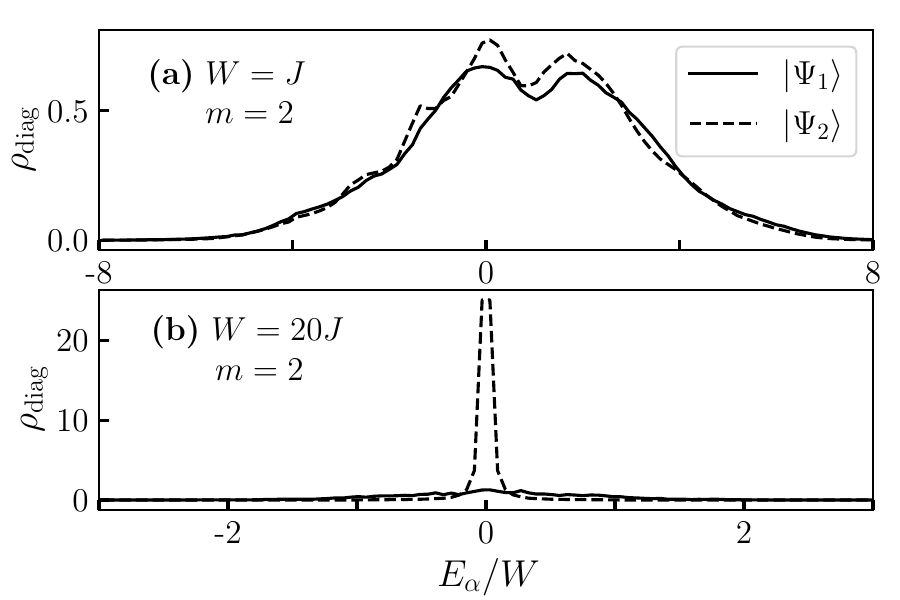}
\caption{The energy distribution $\rho_{\rm{diag}}$ (arbitrary units) of the initial states in the eigenbasis plotted for the two initial states $|\Psi_1\rangle$ (solid lines) and $|\Psi_2\rangle$ (dashed lines) for (a) $W=J$ and (b) $W=20J$.  We can see that the initial state $|\Psi_2\rangle$ has high weights around energy $E_{\alpha}/W=0$ for large disorder strength.
}
\label{fig:c0c1}
\end{figure}

The distinct behavior for the two initial states at large $W$ for the $m=2$ case can be related to the eigenstate properties discussed in  Sec.~\ref{sec:static}. We first check where  the states $|\Psi_1\rangle$ and $|\Psi_2\rangle$ 
have their largest weight in the
eigenspectrum of  $\hat{H}_2$.
To that end, we decompose both states in the eigenbasis $|\psi_\alpha\rangle$ of $\hat{H}_2$
\begin{align}
|\psi(t=0)\rangle = \sum_\alpha c_\alpha |\psi_\alpha\rangle \,.
\end{align}
 We plot the energy distribution $\rho_{\rm{diag}}$ (arbitrary units), which is $|c_\alpha|^2$ multiplied with the density of states (DoS), for the two initial states versus eigenenergy $E_\alpha$ in Fig.~\ref{fig:c0c1}. The DoS is calculated by dividing the spectrum into $100$ bins and counting the number of eigenstates in each bin.
There are no qualitative difference between the two initial states for $W=J$. At large $W=20J$,  $|\Psi_2\rangle$ has a large probability at energy $E_\alpha=0$ in contrast to $|\Psi_1\rangle$, which has contributions from across the spectrum. From the analysis of the entanglement entropy [Fig.~\ref{fig:entr_hist}(d)] and the correlations $C_{2i+1,2i+3}^c$ [Fig.~\ref{fig:eig_corr}(f)], we know  that both quantities  are large around $E_\alpha=0$. These large-entanglement eigenstates and large correlations between clean sites explain the small value of $O_{CDW}$ after a long time for the initial state $|\Psi_2\rangle$. 
The emergence of highly-entangled eigenstates also leads to a much faster time-dependent increase of the entanglement entropy, discussed in App.~\ref{app:svn-time}.

\section{Diagonal ensemble average}\label{sec:diag}
After analyzing the properties of the many-body eigenstates and the temporal dynamics of different initial states, we discuss the infinite-time values of different observables in this section. The diagonal ensemble average determines the infinite-time expectation value of an observable $\hat O$~\cite{Rigol2008}, given by
\begin{equation}
    O_{DE} = \sum_\alpha |c_\alpha|^2 O_{\alpha \alpha},
\end{equation}
where $O_{\alpha\alpha} = \langle \psi_\alpha | \hat O | \psi_\alpha \rangle$ with $|\psi_\alpha\rangle$ being the eigenstates of the Hamiltonian.

\begin{figure}[t]
\centering
\includegraphics[width=1\linewidth]{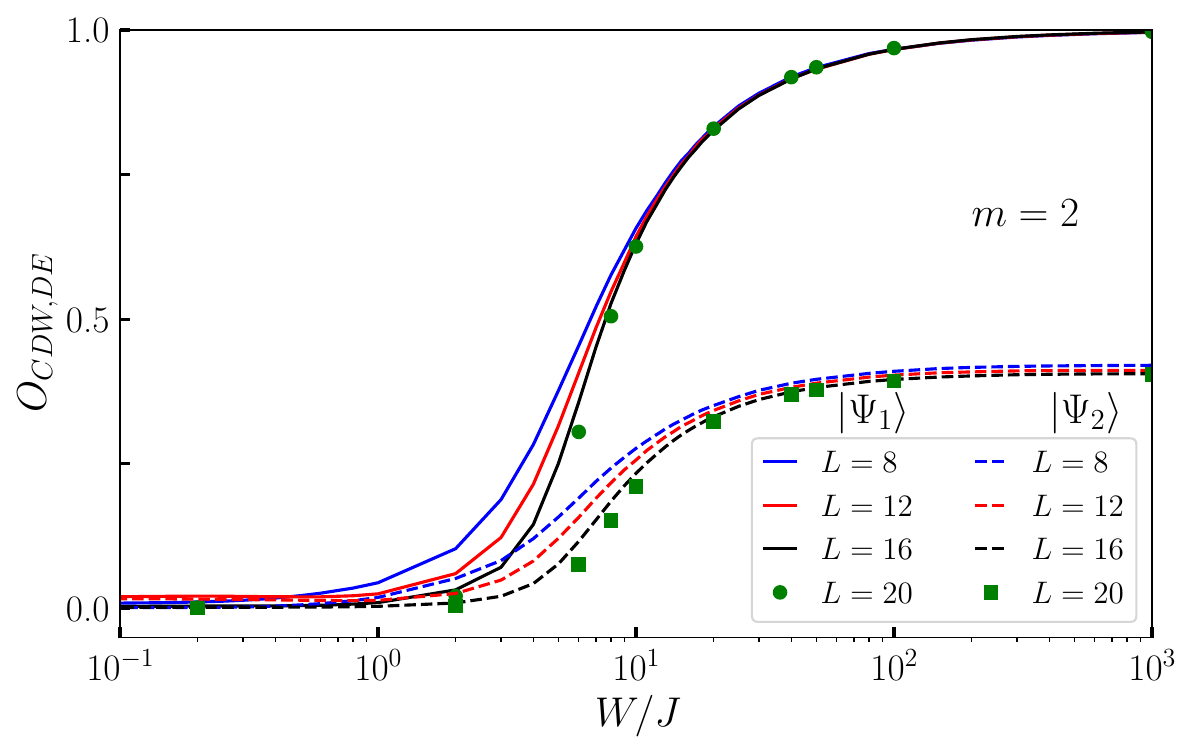}
\caption{Diagonal-ensemble average $O_{CDW,DE}$ of the imbalance  
as a function of disorder strength, computed for the 
two initial states $|\Psi_1\rangle$ and $|\Psi_2\rangle$ represented by two different line styles and for different system sizes $L=8,12,16$.  At large $W$, $O_{CDW,DE}$ saturates at two different values for the two initial states.
% Inset: Extrapolation of the data using a second-order polynomial in $1/L$ for $W=8J$.
}
\label{fig:Ocdwde}
\end{figure}

First, we calculate the diagonal ensemble average of the CDW order parameter ($O_{CDW,DE}$) for the two initial states considered in Sec.~\ref{sec:dynamics} and plot it in Fig.~\ref{fig:Ocdwde} for different system sizes with $m=2$, and over many decades of disorder strength. A drastic difference between the two initial states can immediately be spotted. At large $W$,  $|\Psi_1\rangle $ leads towards complete localization ($O_{CDW,DE} \sim 1$) and $O_{CDW,DE}$ becomes $L$-independent on the system sizes considered here. In contrast, the dynamics starting from $|\Psi_2\rangle $ does not completely localize even for very large $W$ but saturates to a much smaller value of $O_{CDW, DE}$. In this case, we also find a large system-size dependence of $O_{CDW,DE}$, decreasing with increasing $L$, even at huge values of $W$. 
% Moreover, we illustrate  the finite-size dependence  of $O_{CDW,DE}$ at an intermediate value of $W=8J$, shown in the inset of Fig.~\ref{fig:Ocdwde}. The (ad-hoc) second-order polynomial fitted to the finite-size data suggest a much faster delocalization of the $|\Psi_2\rangle$ compared to  the $|\Psi_1\rangle$ on the available system sizes, consistent with a vanishing long-time limit in the thermodynamic limit for $|\Psi_2\rangle$.

Next, we calculate the diagonal ensemble average of the connected  density-density correlations $C_{ij}^c$ (Eq.~\eqref{eq:corr}) to capture the underlying physics for such a significant initial-condition dependent dynamics. 
We calculate three such correlations: nearest-neighbor correlations ($C_{i,i+1,DE}^c$), shortest-distance correlations between disordered sites  ($C_{2i,2i+2,DE}^c$), and shortest-distance  correlations between clean sites ($C_{2i+1,2i+3,DE}^c$) and average over the whole system.

\begin{figure}[t]
\centering
\includegraphics[width=1\linewidth]{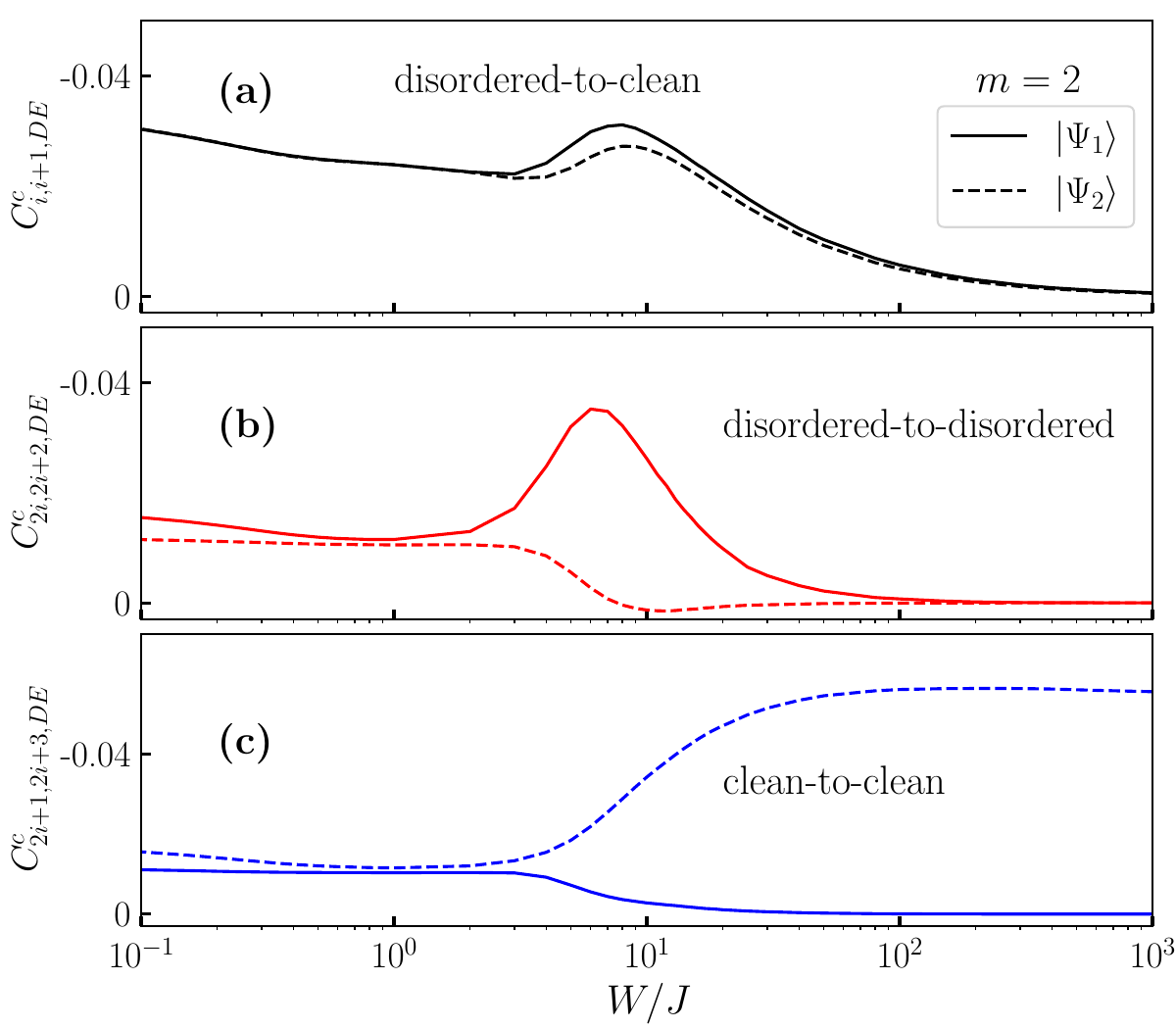}
\caption{Diagonal-ensemble average of correlators $C_{i,j,DE}$ 
as a function of disorder strength, computed for the two 
 different initial states $|\Psi_1\rangle$ and $|\Psi_2\rangle$ represented by two different line styles. We show results for (a) nearest-neighbor correlators $C_{i,i+1,DE}^c$, (b) shortest-distance correlations  $C_{2i,2i+2,DE}^c$ between disordered sites (i.e., $i$ even), and (c) shortest-distance correlators $C_{2i+1,2i+3,DE}^c$  between clean sites (i.e., $i$ odd), represented by black, red and blue lines, respectively. The large non-vanishing values of $C_{2i+1,2i+3,DE}^c$ for the  initial state $|\Psi_2\rangle$ at large $W$ compliment the small values of $O_{CDW,DE}$ seen in  Fig.~\ref{fig:Ocdwde} for the same parameters. This phenomenon is illustrated in Fig.~\ref{fig:model}(b), showing that dynamics between clean sites are possible at large $W$ via resonances while processes between disordered sites are suppressed.
 }
\label{fig:Cde}
\end{figure}

If we compare the two initial states, we see a qualitatively similar behavior at small $W$ values. The system generates correlations, which leads to thermalization. At large $W$, however, for $|\Psi_1\rangle $, all correlations start to vanish, some earlier and some later. In contrast, for the initial state $|\Psi_2\rangle$,  other than $C_{2i+1,2i+3,DE}^c$, which corresponds to the correlation between clean sites, all the correlations vanish at very large $W$. The non-vanishing $C_{2i+1,2i+3,DE}^c$ for $|\Psi_2\rangle$ (see Fig.~\ref{fig:Cde}) at large $W$ explains the tendency for delocalization seen in Fig.~\ref{fig:Ocdwde} for the second initial state.

\section{Conclusions}\label{sec:con}
In conclusion, we studied the delocalization properties of a partially disordered one-dimensional system and compared it with the fully disordered case. In the single-particle spectrum,  extended states exist for the partially disordered case, which do not disappear in the large-disorder regime. 

In the many-body case, the distribution of the half-chain entanglement entropy displays a tail of highly-entangled states, absent in a fully localized system. These states result from resonances
between particles in the clean sites.
This is also reflected in  the density-density correlators between clean sites which remain large with increasing disorder whereas the correlation between clean-to-disordered and disordered-to-disordered sites decreases rapidly. As a result, it should be possible to derive an effective model consisting solely of clean sites in the large disorder limit. The strength of effective hopping matrix elements between the neighboring clean sites is then determined by the second-order hopping process through a disordered site of the original model. This results in an effective model that bears resemblance to a tight-binding random hopping model~\cite{Eggarter1978,Soukoulis1981,Evangelou1986,Roman1987,Roman1988,Akshay2021}. An analysis of such a model is left for future work.

As a consequence of the highly-entangled states, a significant initial state-dependent dynamical behavior is observed in the partially disordered system, in particular, at large disorder. Two representative charge-density-wave states evolve to two very distinct steady states depending on whether the particles are initially localized on disordered or clean sites. Whenever particles originate from clean sites, the residual long-time values are reduced compared to particles spreading out from disordered sites and decay with system size.

Our set-up could be realized in optical-lattice experiments
and would yield an avenue for systematic studies of 
delocalization in initial state-dependent schemes.
Both the initial states and the disorder patterns can be realized using digital mirror devices. We stress the similarity to inverted quantum scars discussed in \cite{Iversen2022,Srivatsa2023,Chen2023}. A natural extension of our work
would be the investigation of patterned disorder in the Bose-Hubbard model realized in some quantum-gas experiments \cite{Choi2016,Leonard2023}.\\

The research data shown in the figures will be made available 
as ancillary files on arXiv.org post publication.

\section*{Acknowledgement}
We acknowledge fruitful discussions with  K. Hazzard, M. Hopjan,  D. Luitz, R. Melko, T. Mishra,  M. Rigol, M. Serbyn, A. \v{S}trkalj, L. Vidmar,  and J. Wang. We thank an anonymous Referee for pointing out the connection between the extension of the
ergodic regime and the width of the disorder distribution in the 
$m>1$ case.
This work was supported by the Deutsche Forschungsgemeinschaft (DFG, German Research Foundation) – 499180199, 436382789, 493420525 via FOR 5522 and large-equipment grants (GOEGrid cluster). This research was supported in part by the National Science Foundation under Grant No. NSF PHY-1748958. F.H.-M. is grateful for the hospitality at KITP, UC Santa Barbara, where part of this work was performed.     This work used the Scientific Compute Cluster at GWDG, the joint data center of Max Planck Society for the Advancement of Science (MPG) and University of G\"ottingen.

\section{Appendix}\label{sec:app}

\begin{figure}
\centering
\includegraphics[width=1\linewidth]{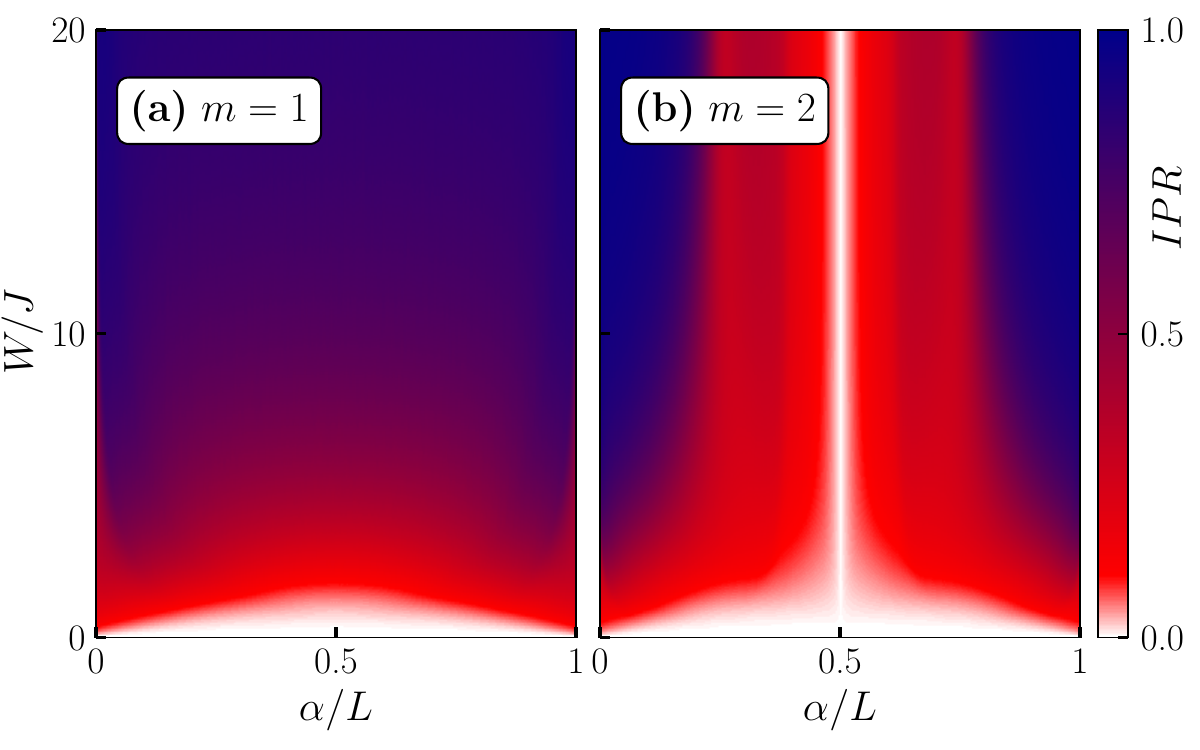}
\caption{Non-interacting case: Existence of  extended states at large disorder strength in the single-particle spectrum of a partially disordered system. The $IPR$ of all the single-particle eigenstates $|\psi_\alpha\rangle$ is plotted with respect to the disorder strength for (a) $m=1$  and (b) $m=2$. Here, we consider $L=3000$ and the $IPR$ is averaged over $N_r = 1000$ realizations. The $m=2$ case exhibits a region with   $IPR\sim 0$  at the center of the spectrum, signifying the existence of an extended state in the partially-disordered case.}
\label{fig:ipr}
\end{figure}

\subsection{Noninteracting case}
Here, we want to discuss the single-particle physics of the partially disordered system and compare it to the Anderson disorder model in this section. $\hat{H}_{m=1}$ with $V=0$ is the well-known Anderson disorder model, where all the states are localized for $W/J>0$. In contrast, the system exhibits extended states for $m>1$ in the spectrum. We calculate the inverse participation ratio ($IPR$), a marker to distinguish extended and localized states, for $m=1$ and $m=2$ to depict the physics. The $IPR$ of an eigenstate $|\psi_\alpha\rangle$ can be calculated as,
\begin{equation}
    IPR_\alpha = \sum_{l=1}^L |\psi_{\alpha, l}|^4
\end{equation}
where $l$ is the site index and $\psi_{\alpha, l}$ is the amplitude of $|\psi_\alpha\rangle$ on site $l$. The $IPR_\alpha$ is proportional to $1/L$ for an extended state and is $\mathcal{O}(1)$ for localized states. In Fig.~\ref{fig:ipr}(a) and (b), we show $IPR_\alpha$ in the spectrum with varying $W$ for $m=1$ and $m=2$, respectively. While Fig.~\ref{fig:ipr}(a) shows trivial Anderson localization of the eigenstates, Fig.~\ref{fig:ipr}(b) clearly shows the existence of an extended state at the middle of the spectrum for all disorder strength $W$.

One can show from an elementary analysis that, for the non-interacting $\hat{H}_{m=2}$, the system can have one special periodic eigenstate $|\psi_p\rangle = \sum_{l\in[1,L,4]} (|l+1\rangle - |l+3\rangle)$ with eigenenergy $E_p=0$. Similarly for a non-interacting $\hat{H}_{m=3}$ there exist two periodic states
with $E_p^\pm = \pm J$ and so on. 
These states can survive and remain delocalized for any $W>0$. In general, we find $m-1$ number of exactly periodic states in the single-particle spectrum of the Hamiltonian $\hat{H}_m$ (not shown). 

\subsection{Finite-size dependence of eigenstate entanglement entropy}
\label{app:svn-L}

\begin{figure}[t!]
\centering
\includegraphics[width=1\linewidth]{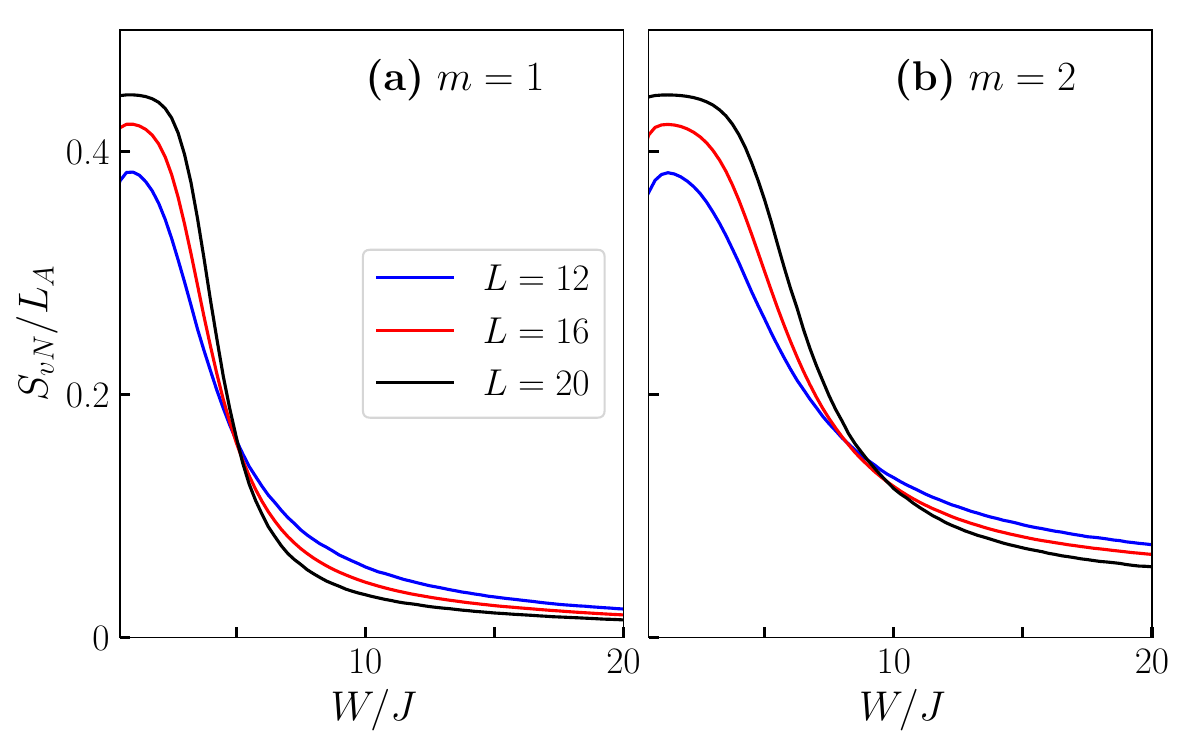}
\caption{Von Neumann entanglement entropy $S_{vN}$ at energy density $\epsilon = 0.5$ shown for different system sizes and for the same the parameters as in (a) Fig.~\ref{fig:pds}(a) and (b) Fig.~\ref{fig:pds}(b), respectively.}
\label{fig:entr}
\end{figure}

The finite-size trends in  $S_{vN}$ are captured in Fig.~\ref{fig:entr}. Here, we plot  $S_{vN}/L_A$, averaged over the states present in the middle bin corresponding to $\epsilon = 0.5$ of Fig.~\ref{fig:pds}. Note that for $L=20$, we use the shift-and-invert method to find fifty eigenstates around energy density $\epsilon = 0.5$ and calculate the averaged $S_{vN}/L_A$. The $m=1$ and $2$ cases look qualitatively similar, but the entropy at large $W$ is larger for $m=2$ compared to the corresponding $m=1$ cases. 

\subsection{Time-dependence of the entanglement entropy}
\label{app:svn-time}

\begin{figure}[t]
\centering
\includegraphics[width=1\linewidth]{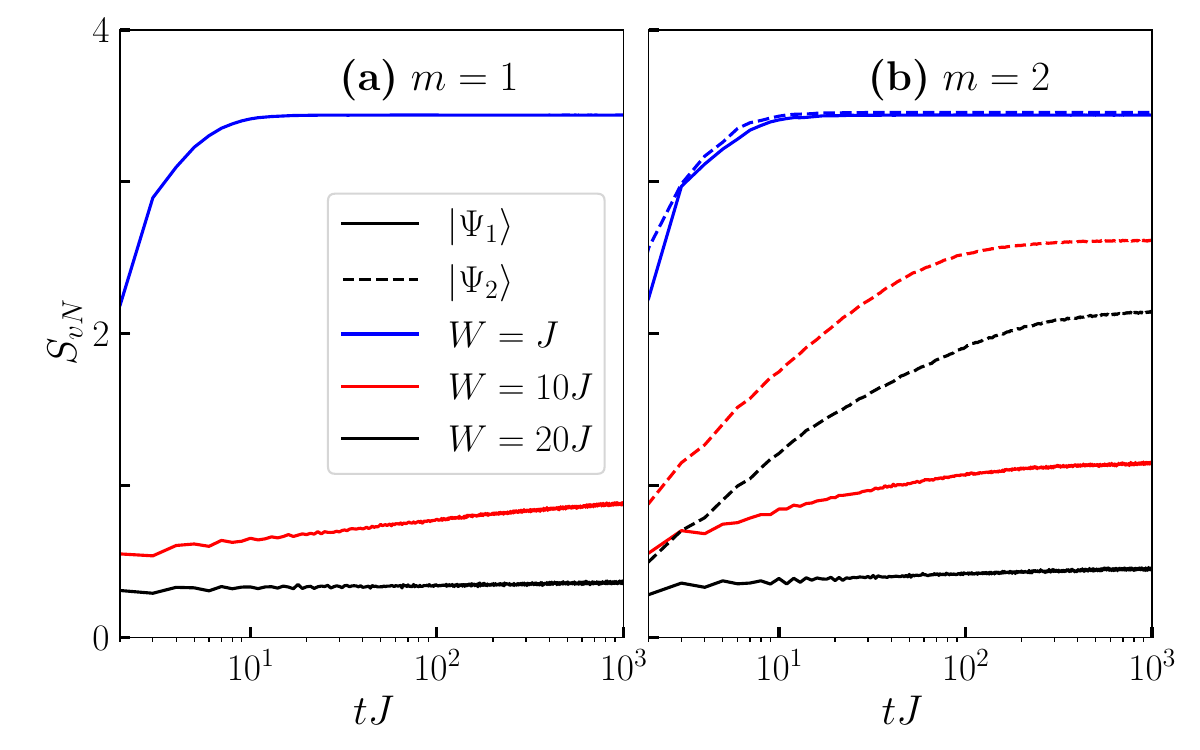}
\caption{Time evolution of the entanglement entropy $S_{vN}$ for (a) $m=1$ and (b) $m=2$  and $L=16$ sites.  Different values of $W$ are represented by different colours and data for the two initial states $|\Psi_1\rangle $ (Eq.~\eqref{eq:psi1}) and $|\Psi_2\rangle$ (Eq.~\eqref{eq:psi2})  are represented by the solid and dashed lines, respectively. In both the $m=1$ and $2$ case and for small $W=J$, $S_{vN}$ quickly reaches a large value and saturates, while  for large $W=20J$, we see a logarithmic increase. For the $m=2$ case and for large disorder strength,  $S_{vN}$ increases faster for the initial state $|\Psi_2 \rangle$  compared 
to the state $|\Psi_2\rangle$.}
\label{fig:entrt}
\end{figure}

We also calculate and analyze the evolution of entanglement entropy $S_{vN}$ to complement the aforementioned finding. We plot the time-dependent $S_{vN}(t)$ in Figs.~\ref{fig:entrt}(a) and (b) for the initial states and model parameters considered in Fig.~\ref{fig:Ocdwt}(a) and (b), respectively. The logarithmic growth of $S_{vN}$ for $m=1$ case with large $W=10J$ and $20J$ indicates typical behavior of the MBL phase \cite{Bardarson2012,Znidaric2008}, and the rapid growth of the same for small $W=J$ indicates the thermalization of the system. For the $m=2$ case, we see a similar behavior for smaller $W=J$ since the system thermalizes in this parameter regime. At large $W=10J$ and $20J$, however, there is a clear distinction between the two initial states. In the intermediate time dynamics, $S_{vN}$ grows as a logarithm, implying MBL-like behavior for both  initial states. However, $S_{vN}$ for the initial state $|\Psi_2\rangle$ increases comparatively faster than for $|\Psi_1\rangle$  and saturates at a higher value after a long time.

\begin{figure}[t]
\centering
\includegraphics[clip, trim={{0.0\linewidth} {0\linewidth} {0.0\linewidth} {0.88\linewidth}}, width=1\linewidth]{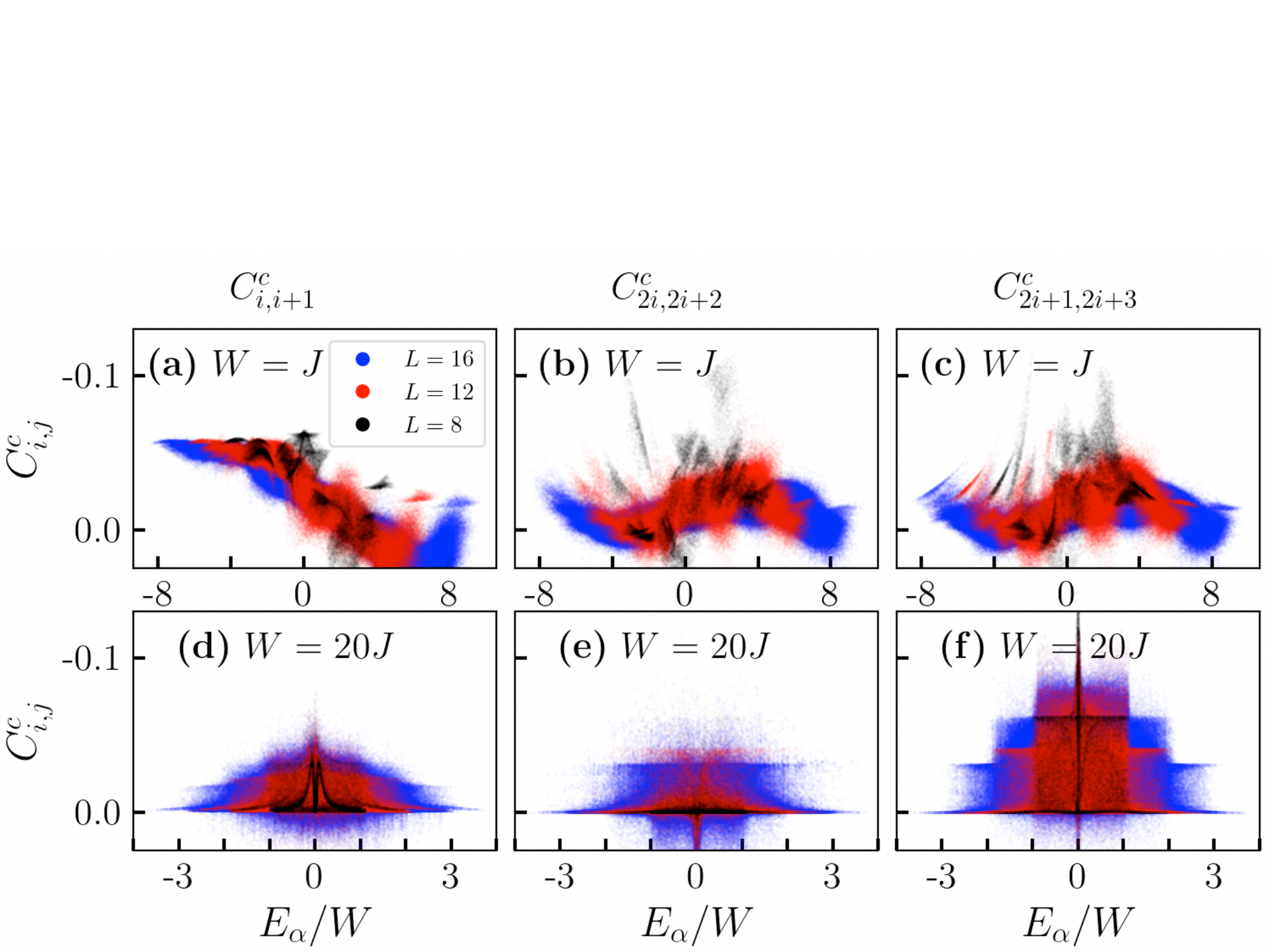}
\caption{System-size dependence of the nearest-neighbor correlations between disordered-to-clean ($C_{i,i+1}^c$), disordered-to-disordered ($C_{2i,2i+2}^c$), and clean-to-clean ($C_{2i,2i+3}^c$) sites. For small $W=J$, the distribution becomes smoother as a function of energy with increasing $L$. For large $W=20J$, the step-like structure for $C_{2i,2i+3}^c$ emerges with increasing system size.}
\label{fig:eig_corr1}
\end{figure}

\subsection{Finite-size dependence of density correlators}
\label{app:corr-C}
Here, we explore the finite-size dependence of the density correlations discussed in Sec.~\ref{sec:static}. We plot the correlations $C_{i,j}^c$ at small ($W=J$) and large disorder strengths ($W=20J$) in Fig.~\ref{fig:eig_corr1}(a-c) and Fig.~\ref{fig:eig_corr1}(d-f), respectively, for the $m=2$ case. For $W=J$, the correlations become smoother function of energy for larger $L$. Also, the distribution of  $C_{i,j}^c$ shrinks at a given energy which is expected for an ergodic system obeying ETH. In the large $W$ case, we see non-ETH-like behavior and the wedding-cake structure emerges with increasing system size on the 
available system sizes.

\subsection{Results for the $\rho = 1/2$ case}
\label{app:corr-D}

\begin{figure}[t]
\centering
\includegraphics[width=1\linewidth]{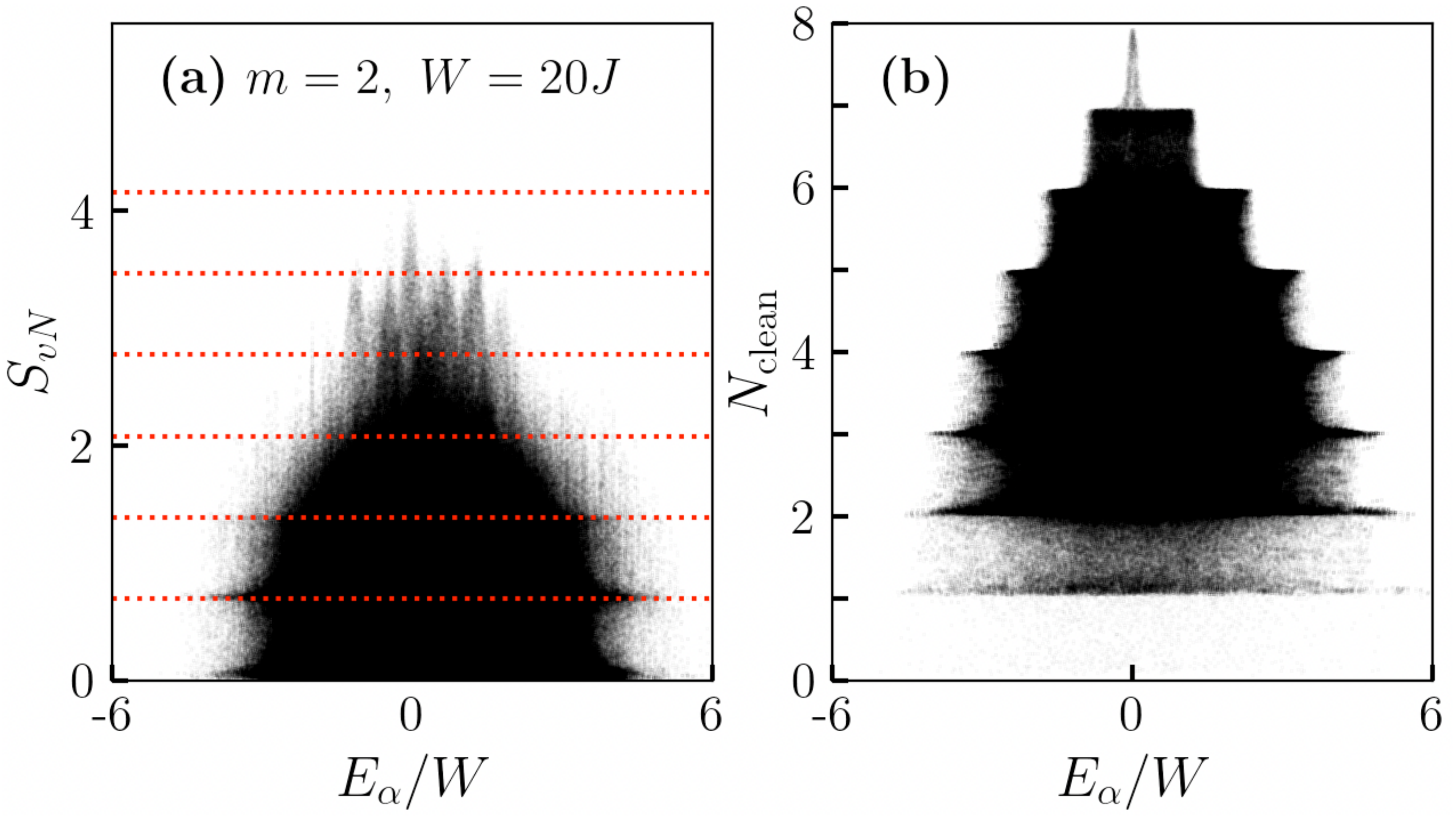}
\caption{(a) Entanglement entropy and (b) number of particles in the clean sites versus eigenstate energy for $m=2$ case with $L=16$, $\rho =1/2$, and $W=20J$. The red dotted lines in (a) are integer multiples of ${\rm{ln}}(2)$.}
\label{fig:entr_nb2}
\end{figure}

For our main results presented above, we have considered the filling of $\rho = 1/4$. However, it is important to note that the physics we are discussing is not limited to this specific filling. In this section, we present our findings for a filling of $\rho=1/2$. Similar to Fig.~\ref{fig:entr_nb}(e) and (f), we have plotted the entanglement entropy and the number of particles in clean sites for the $m=2$ case with $L=16$, $\rho=1/2$, and $W=20J$ in Fig.~\ref{fig:entr_nb2}(a) and (b), respectively. We can clearly observe the highly-entangled states and the wedding cake-like structure of $N_{\rm{clean}}$ in this case as well. However, unlike in the $\rho =1/4$ case where the highest entangled eigenstates correspond to the states with $N_{\rm{clean}} \sim N$, for $\rho = 1/2$, eigenstates with $N_{\rm{clean}} \sim N$ are (trivially) localized states.  
For the $\rho = 1/2$ case, if $N_{\rm{clean}} = N$, all the clean sites are filled leading to an energy $E \sim 0$, and the sites in the neighbors (disordered sites) are not correlated at large $W$, which explains the localized nature of these eigenstates. The large entangled states visible in Fig.~\ref{fig:entr_nb2}(a) mostly correspond to the case where the clean sites are half filled, i.e., $N_{\rm{clean}} = N/2$.
Therefore, the filling $\rho=1/4$ chosen in the main text 
gives a more illustrative example of the connection between $S_{vN}$ and $N_{clean}$.
We can also capture this aspect in the time evolution of different initial states discussed below. There, we find that the initial state with $N_{\rm{clean}} = N/2$ relaxes significantly compared to the initial states with $N_{\rm{clean}} = N$ and $N_{\rm{clean}} = 0$.

\begin{figure}[t!]
\centering
\includegraphics[width=1\linewidth]{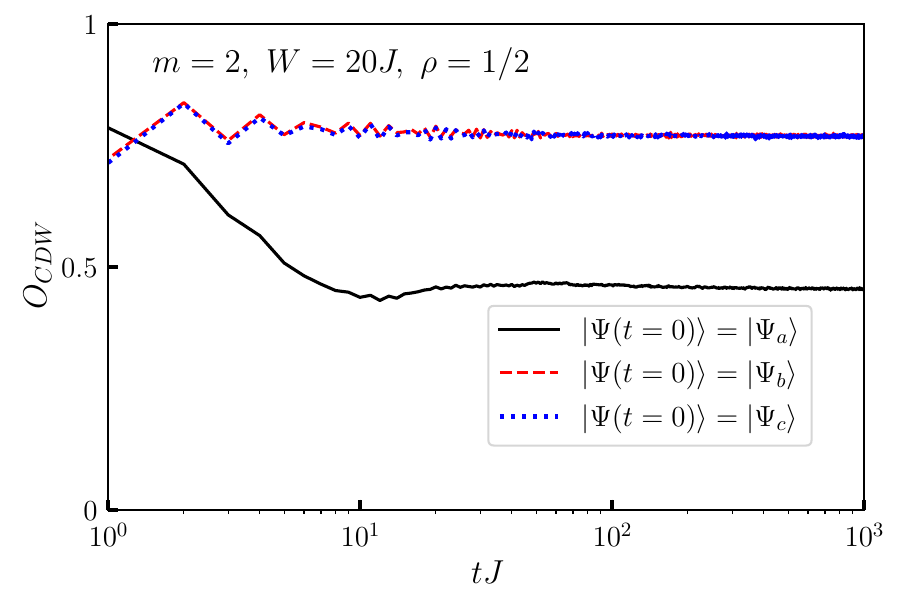}
\caption{Time evolution of the CDW order parameter corresponding to the three initial states $|\Psi_a\rangle$, $|\Psi_b\rangle$ and $|\Psi_c\rangle$  shown for the $m=2$ case with $L=16$, $\rho =1/2$ and $W=20J$.}
\label{fig:It}
\end{figure}

As mentioned above, we also see the initial-state dependent dynamics for $\rho =1/2$ filling for $m=2$ case. Here, we consider three initial states:
\begin{equation}
    |\Psi_a\rangle = \prod_{i=0}^{L/4-1} \hat b^\dagger_{4i+1} \hat b^\dagger_{4i+2} |0\rangle
\end{equation}
where half of the particles are in clean sites and the other half in disordered sites,
\begin{equation}
    |\Psi_b\rangle = \prod_{i=0}^{L/4-1} \hat b^\dagger_{4i+1} \hat b^\dagger_{4i+3} |0\rangle
\end{equation}
where all particles are in disordered sites, and 
\begin{equation}
    |\Psi_c\rangle = \prod_{i=0}^{L/4-1} \hat b^\dagger_{4i+2} \hat b^\dagger_{4i+4} |0\rangle
\end{equation}
where all particles are in the clean sites. The initial states are time evolved and the order parameter for this initial state, given by
\begin{equation}
    O_{CDW} = \frac{N_A - N_B}{N}\,,
\end{equation}
is monitored. 
This $O_{CDW}$ is identical to the imbalance in the density between initially occupied and unoccupied sites.
Here, $N_A$ ($N_B$) is the total number of particles in initially occupied (unoccupied) sites. As we can see from the Fig.~\ref{fig:It}, the initial state $|\Psi_a\rangle$ shows different relaxation dynamics (similar to $|\Psi_2\rangle$ in $\rho=1/4$ case) compared to the other two initial states.

\bibliography{references}

\end{document}